\documentclass{pasa}

\title[MGPS--2 Galactic Plane Survey]{The second epoch Molonglo Galactic Plane Survey: images and candidate supernova remnants}
\author[A. J. Green et al.]{A. J. Green$^{1}$, S. N. Reeves$^{1}$, T. Murphy$^{1}$ 
\thanks{Corresponding author. Email: anne.green@sydney.edu.au}\\
\affil{$^1$ Sydney Institute for Astronomy, School of Physics, The University of Sydney, NSW 2006, Australia}}%

\jid{PASA}
\doi{10.1017/pasa\the\year.37}
\jyear{\the\year}

\newcommand{\arcsec}{^{\prime\prime}}
\newcommand{\arcmin}{^{\prime}}

\newcommand{\aap}{A\&A}

\newcommand{\aj}{AJ}
\newcommand{\apj}{ApJ}

\newcommand{\apjs}{ApJS}
\newcommand{\mnras}{MNRAS}
\newcommand{\pasa}{PASA}

\begin{document}%
\begin{abstract}
The second epoch Molonglo Galactic Plane Survey (MGPS-2) covers the area $245^{\circ} \leq l \leq 365^{\circ}$ and $|b| \leq 10^{\circ}$ at a frequency of 843 MHz and an angular resolution of $45\arcsec\times45\arcsec$cosec($\delta)$. The sensitivity varies between $1 - 2$ mJy beam$^{-1}$ depending on the presence of strong extended sources. This survey is currently the highest resolution  and most sensitive large-scale continuum survey of the southern Galactic Plane. In this paper, we present the images of the complete survey, including postage stamps of some new supernova remnant (SNR) candidates and a discussion of the highly structured features detected in the interstellar medium. The intersection of these two types of features is discussed in the context of the``missing"  SNR population in the Galaxy. 
\end{abstract}
\begin{keywords}
Milky Way Galaxy -- interstellar medium -- supernova remnants -- radio continuum -- surveys
\end{keywords}
\maketitle%
\section{INTRODUCTION}
\label{sec:intro}
To fully understand the complex processes and ecosystems which influence the structure and evolution of the Galaxy, it is essential to have panoramic surveys of high angular resolution and sensitivity. 
The Molonglo Observatory Synthesis Telescope (MOST; Mills 1981, Mills 1985, Robertson 1991) has at present the largest collecting area of cm-wavelength radio telescopes in the southern  hemisphere. With sub-arcmin resolution and brightness sensitivity of $\sim1$~K it is very well suited to imaging the Galactic plane which has radio emission detectable on a wide range of angular and intensity scales. 

The second epoch Molonglo Galactic Plane Survey (MGPS-2) was undertaken concurrently with the Sydney University Molonglo Sky Survey (SUMSS; Bock, Large \& Sadler 1999, Mauch et al. 2003) to produce deep imaging coverage of the whole sky south of  
$\delta = -30^{\circ} $, which is the northern limit of MOST for a fully synthesized image. The scale and detail provided by MGPS-2 makes it a valuable asset in undertaking multwavelength studies of the interstellar medium (ISM).

MGPS-2 covers an area of 2400  deg$^2$ defined as  $245^{\circ}\leq l \leq 365^{\circ}$ with $|b| \leq 10^{\circ}$  in radio continum at 843~MHz with an angular resolution of  $45\arcsec\times45\arcsec$cosec($\delta$). The overarching goal of MGPS-2 was to study the processes and structures in the ISM and how they might increase our understanding of the star formation cycle. Murphy et al. (2007) provided a catalogue and analysis of the compact sources from the survey, and Bannister et al. (2011a, 2011b) investigated variability in source intensities using MGPS-2 and archival MOST data. 

This paper presents the complete set of images from MGPS-2 with the goals of identifying a sample of new supernova remnants (SNRs) in the diffuse extended emission seen in the survey and investigating the large filamentary structures which may not be discrete sources but part of the larger Galactic scaffolding. Representative images of the survey are shown here and the complete set of mosaics can be found at www.physics.usyd.edu.au/sifa/Main/MGPS2.

In Section 2, we review the observations and data products from MGPS-2. Section 3 presents the new SNR candidates  and the search criteria used. Section 4 describes the extended filamentary structures which appear to be predominantly thermal and a property of the ISM rather than discrete sources. In Section 5 we discuss the implications of our survey in explaining the SNR population and the ecology of the ISM and the value of the survey as a resource to anchor future observations from southern low frequency instruments such as the Murchison Widefield Array (MWA; Bowman et al. 2013,Tingay et al. 2013). A summary of our conclusions are given in Section 6. 

\section{MGPS-2 SURVEY STRATEGY \& DATA PRODUCTS}
The MOST is an east-west synthesis interferometer, with cylindrical geometry and continous $uv$-coverage from baselines of 16 m to 1.6 km. The telescope produces a fully synthesised image of an area 
$>5.5$ deg$^{2}$ in 12 hours for Declinations $\leq-30^{\circ}$. The MGPS-2 is a survey of the Galactic Plane undertaken concurrently with the extragalactic SUMSS survey over the period 1997 to 2006.  The technical specifications of the telescope and the survey parameters are given in Table 1. 

Because of the complexity and strength of the emission from the Galactic Plane the data quality is dynamic range limited and the strategy for observations was to complete the coverage expeditiously since achieving uniform sensitivity was not possible. The data are not noise limited. The confusion limit for the telescope, with its given frequency and angular resolution, is 120~$\mu$Jy beam$^{-1}$, well below the $1\sigma$ rms achieved for the survey. 

The pointing grid for observations is shown in Figure 1, taken from Murphy et al. (2007).  The sensitivity varies with declination, the relative position in the field of view and the operational performance of the telescope. The telescope forms fan beams on the sky which are then accumulated over the course of an  observation. It does not record $uv$-visibilities and hence, the calibration of individual baselines is not possible. However, the telescope elements are combined so that most baselines are observed many times, making the array highly redundant. 

The calibration of the telescope for both flux density and position is described in Bock et al. (1999). The calibration is based on a sample of strong unresolved sources measured by Campbell-Wilson \& Hunstead (1994). The shortest baseline that the telescope records is 16m, which means that smooth structures with a scale larger than $\sim25^{\prime}$ are resolved out and the total fluxes of large sources are underestimated. However, the telescope is very sensitive to filaments and edges. There is no opportunity at present to add single dish observations to enable total flux estimates.

\begin{figure}
\begin{center}
\includegraphics[width=\columnwidth]{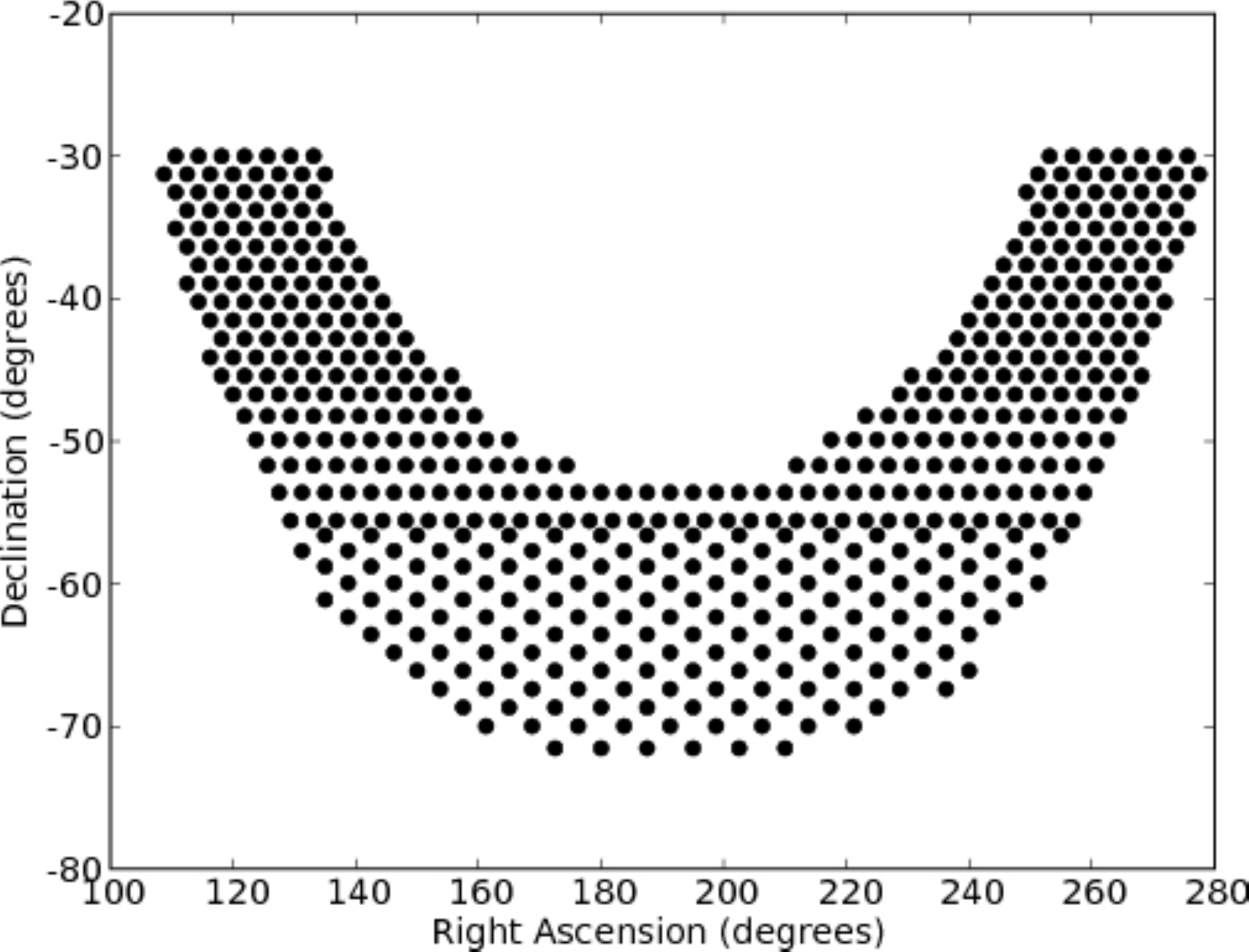}
\caption{The hexagonal pointing centre grid for MGPS-2, selected to minimize the time for survey coverage. Figure reproduced from Murphy et al. (2007).}\label{mgps-grid}
\end{center}
\end{figure}

After calibration and quality control, the individual images were combined into mosaics $4.3^{\circ}\times4.3^{\circ}$, which are slightly overlapped for continuity in analysing large-scale structure. The beam for each mosaic was held constant and was calculated for its central position, which means the southern section of each mosaic has a beam slightly broader than the natural data resolution and the northern part is slightly over-resolved.  The variation is minor for the Declination range covered by MGPS-2 and was done to enable quantitative analysis of the mosaics. The pointing centres were chosen to mesh with SUMSS and are in equatorial, rather than Galactic coordinates. This is to preserve a quantifiable beam shape for each mosaic as the natural observing mode of the telescope is in Equatorial.

Two representative mosaics from MGPS-2 are shown. Mosaic J1600M52 (Figure 2) is from a region encompassing  the tangent direction to the spiral arm of the Galaxy near Longitude $330^{\circ}$ (Norma Arm) showing the complex emission from both discrete sources and more extensive filamentary emission. Mosaic J1448M52 (Figure 3) shows a region $\sim7^{\circ}$ away from the Galactic Plane, demonstrating that most of the diffuse emission is now smooth and resolved out by the MOST. In both images there are telescope artefacts seen as grating rings and radial spokes around bright sources. These are a consequence of the telescope geometry and typically only a minimal impediment in extended source searches. The vast majority of the unresolved sources are background galaxies and quasars.

\begin{table*}
\caption{Technical specification of MGPS-2.}
\begin{center}
\begin{tabular*}{\textwidth}{@{}l\x r@{}}
\hline
Parameter & Value \\
\hline%
Telescope Location & Latitude $-35.3708^{\circ}$ S  Longitude 149.249$^{\circ}$ E \\
Telescope geometry &  Two cylindrical paraboloids, aligned EW \\ 
Maximum baseline & 1600 m \\
Minimum baseline (separation between E and W arms) & 16 m \\
Central Frequency & 843 MHz \\
Bandwidth & 3 MHz \\
Polarisation &  Right Hand Circular (IEEE) \\
Angular Resolution & $45\arcsec\times45\arcsec$cosec($\delta$) \\
Field of View (single pointing) &  $163\arcmin \times163\arcmin$cosec($\delta$) \\
Number of fields observed & 621 \\
Sensitivity (1$\sigma$ after 12 hour fully synthesized image) & 1 -- 2 mJy beam$^{-1}$ \\
Surface brightness Sensitivity & 0.9 -- 1.8 K \\
Confusion limit for the MOST & 120 $\mu$Jy beam$^{-1}$ \\
Dynamic Range (limited by strong Galactic sources)  & $\sim100:1$ \\
Area Surveyed & 2400 deg$^2$ \\
Galactic Longitude range & $245^{\circ}\leq l \leq 365^{\circ}$ \\
Galactic Latitude range  & $|b| \leq 10^{\circ}$ \\
Data products (mosaic images)$^a$ & $4.3^{\circ}\times4.3^{\circ}$ \\
\hline\hline  \\
\end{tabular*}
\end{center}
\label{tab1}
$^a$ Published data are the extragalactic survey Sydney University Molonglo Sky Survey (SUMSS; Bock et al. 1999; Mauch et al. 2003) and MGPS-2 Compact Source Catalogue (Murphy et al. 2007).\\

\end{table*}

\begin{figure*}
\begin{center}
\includegraphics[width=17.5cm]{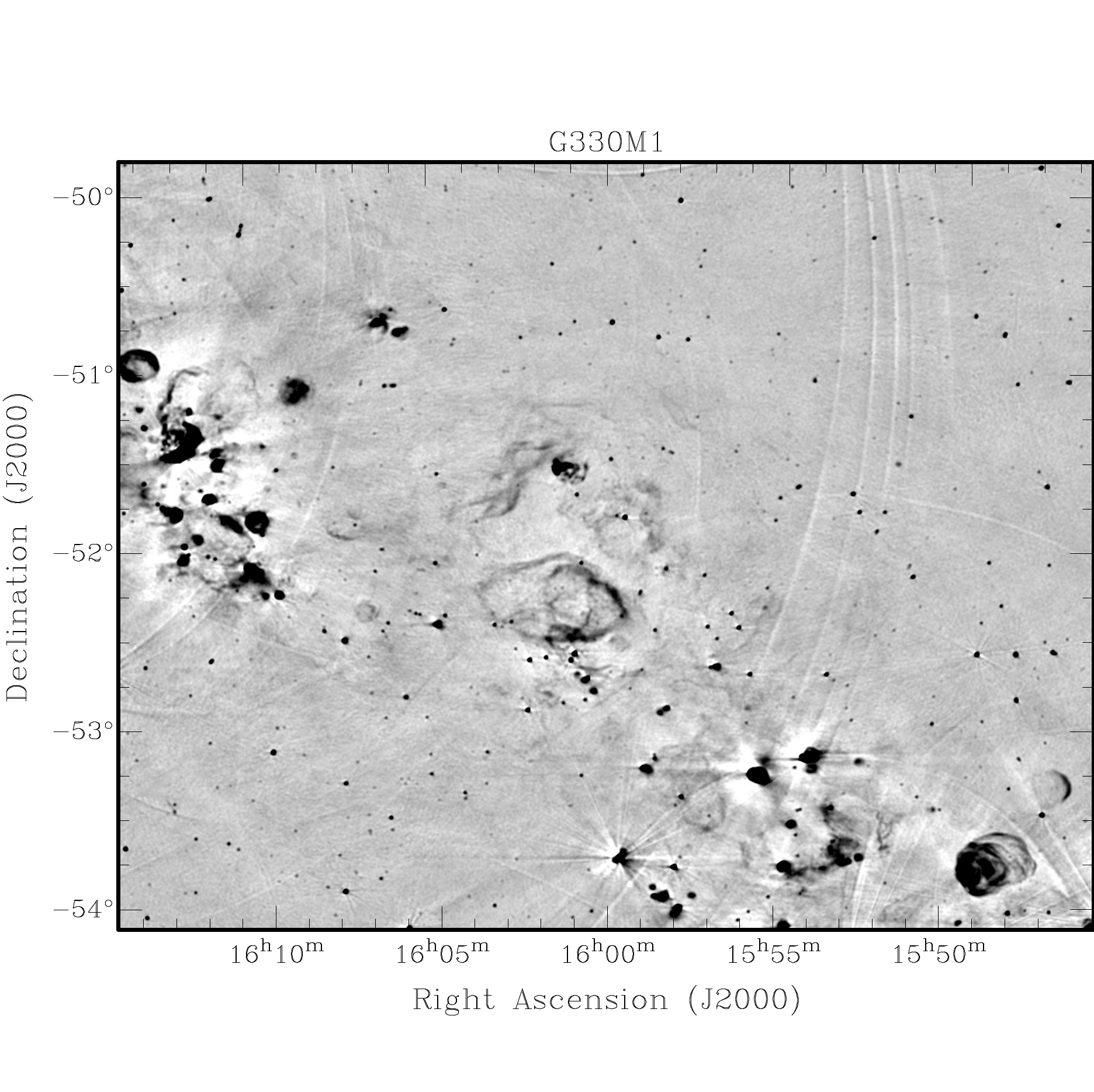}\\
\caption{A representative mosaic from the total of 144 constituting MGPS-2 illustrates the range of intensities seen in the survey and hence the challenge for maintaining constant sensitivity. Mosaic J1600M52 shows a region along the tangent direction of the Norma spiral arm (Galactic Longitude $330^{\circ}$). The greyscale is clipped between $-15$ and $+50$ mJy beam$^{-1}$ to emphasise extended emission. Low level grating artefacts due to out-of-field strong sources can be seen}\label{mgps-mosaic1}
\end{center}
\end{figure*}

\begin{figure*}
\begin{center}
\vspace*{-2cm}
\includegraphics[width=17.5cm]{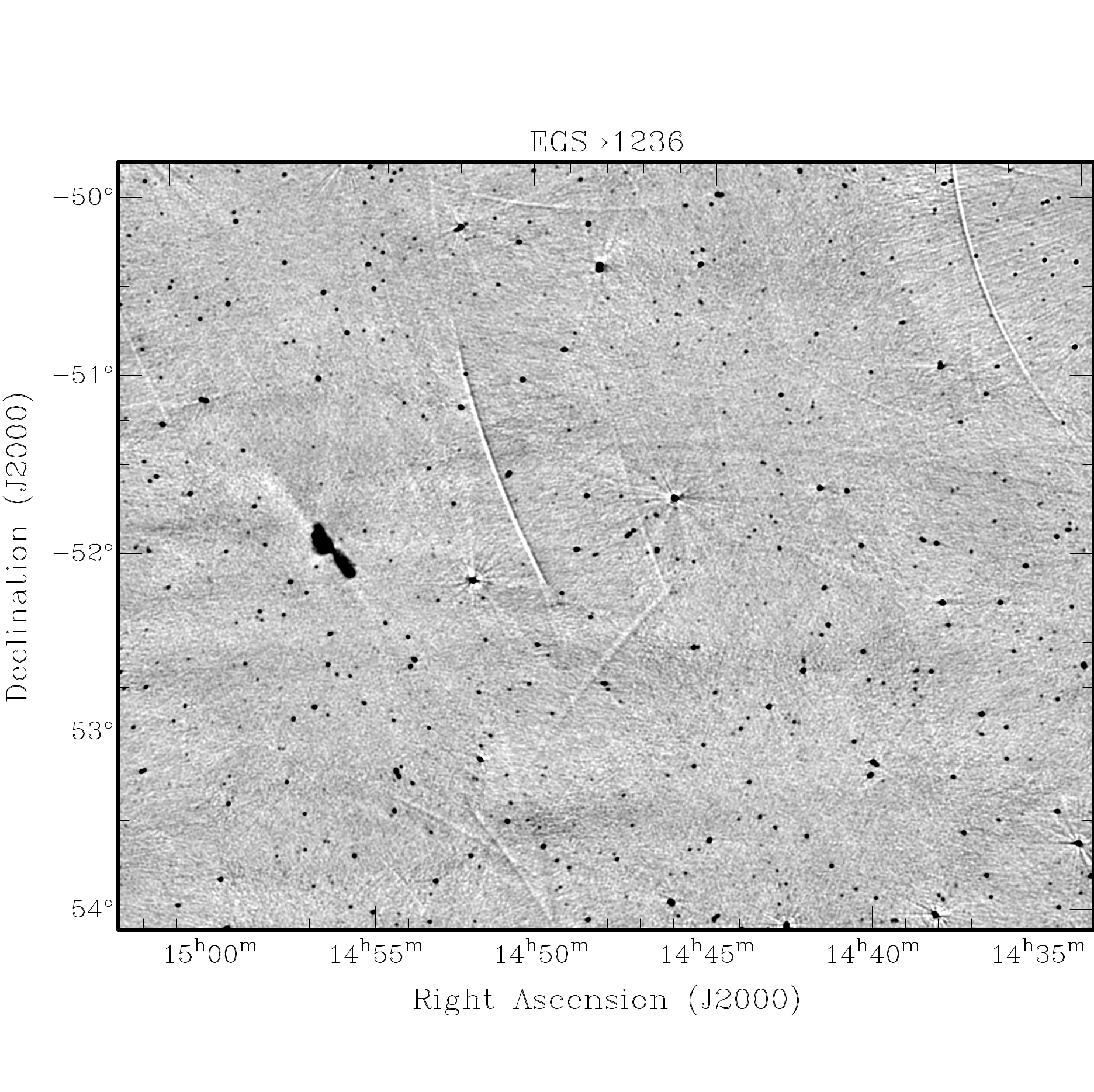}
\caption{A second representative mosaic from MGPS-2, mosaic J1448M52, shows a section of the survey $\sim7^{\circ}$ away from the Galactic Plane demonstrating that the smooth diffuse emission is now largely resolved out by the MOST. A radio galaxy (B1452--517), first imaged by Jones \& McAdam (1992) is seen east of the centre of the mosaic. The greyscale is clipped between $-4$ and $+13$ mJy beam$^{-1}$ to emphasise extended emission. Low level banding in this mosaic is due to solar interference.}\label{mgps-mosaic2}
\end{center}
\end{figure*}

\section{A SAMPLE OF NEW SUPERNOVA REMNANT CANDIDATES}
There is a serious lack of confirmed SNRs in the Galaxy, compared with the expected number extrapolated from external galaxies or from models of Galactic sources and SNR lifetimes. It is predicted that we should see about 1000 SNRs at any one time (Li et al. 1991, Frail et al. 1994, Tammann et al. 1994). Green (2009) records 274 identified objects. Since then there have been several new identifications (Green 2014) but the total number is still well below the predicted value. Only two new SNRs have been published in the central $10^{\circ}$ strip of  MGPS-2 that was searched for SNR candidates and both are below the angular size cutoff set for the present survey (Reynolds et al. 2013, Roy \& Pal 2013) .

SNRs have an average lifetime of $\sim10^5$ years (Frail et al. 1994). Following the SN explosion there is a short period ($\sim1000$ years) of free expansion before the SNR enters the adiabatic Sedov-Taylor phase and interacts with the ISM. There follows a radiative phase (snowplow) and final dissipation. For most of their lifetime radio emission is the principal radiation detected. (e.g. Weiler \& Sramek 1988).

The most compelling evidence for an SNR identification is a negative spectral index, detectable polarisation (from the synchrotron process driving the formation of the radio emission) and the absence of strong mid-infrared (MIR) emission, predominantly produced in the ISM by reradiation from dust grains. 

Large molecules such as polycyclic aromatic hydrocarbons (PAHs) are largely destroyed by the supernova (SN) explosion and evolution, although there is also evidence that some dust is created in the process and produces emission in the ejecta. Rho et al. (2008) found evidence for emission from the ejecta in Cas A by dust heated to 60 -- 120 K and Sibthorpe et al. (2010) reported far-infrared and submillimeter observations suggesting cooler dust, also in Cas A. 

More sensitive and higher resolution surveys of the Galactic Plane at MIR wavelengths have been produced, particularly with the Infrared Array Camera (IRAC) and Multiband Imaging Photometer (MIPS) on the Spitzer Space Telescope. Emission has been detected from roughly 20 -- 30\% of SNRs (Reach et al. 2006, Pinheiro Goncalves et al. 2011). Most of the dust emission comes from interaction of the SNR shock wave with the ISM (e.g. Douvion et al. 2001, Borkowski et al. 2006) and is strongest at 24  and 70 $\mu$m.   Pinheiro Goncalves et al. (2011) notes that the stochastically heated PAH emission at 8$\mu$m is much weaker than at 24$\mu$m, which is produced when SNR shocks interact with the ISM to disrupt large grains and molecules. Previous use of {\it IRAS} ratios of $60/100\mu$m  have been limited by resolution and sensitivity in discriminating SNRs from the diffuse background. 

\subsection{Search Criteria}

The MGPS-2 image database was searched manually to identify new SNR candidates. The principal criteria for selecting new SNRs were the source morphology and the well established anticorrelation between radio emission and 8$\mu$m IR emission (e.g. Whiteoak \& Green 1996, Cohen \& Green 2001, Brogan et al. 2006). 

Sources were selected to be extended sufficiently in the images so that unambiguous shell structure could be observed. The angular resolution of the MOST is $45\arcsec\times45\arcsec$cosec($\delta$) and a minimum diameter of  $\geq 5^{\prime}$ was set. Clearly, our results do not include small diameter SNRs, either young (e.g. Reynolds et al. 2013, Roy \& Pal 2013) or distant.  In addition, the morphology should be shell-like or of a composite morphology, consistent with the evolution expected for SNRs. Where multiwavelength data are available, the source should have a nonthermal spectral index.

\begin{table*}
\caption{Supernova Remnant Candidates from the present survey -- they are not listed in Green (2009). The column headings are described in the text.}
\begin{center}
\begin{tabular}{lccccrrrrr}
\hline
Source & RA & Dec & Size & Peak Flux & Flux & rms & $\Sigma_{843}$ & Fig. \\
~ & (J2000) & (J2000) & (arcmin$^2$) & (mJy/bm) & (Jy) & (mJy/bm) & ($10^{-5}$ Jy/sr) & ~  \\

\hline%

G269.7+0.0 &    09:10:47.7 & $-$48:03:16   & $33\times27$   &   9   &  --            & $>3$  & --      &  4(a)  \\
G291.0+0.1  & 11:12:33.3 & $-$60:28:42 & $22\times20$  & 18   &  0.58  & $>3$ &   0.16   &  4(b) \\
G296.6$-$0.4  & 11:55:50.8 & $-$62:34:27 & $14\times10$  & 18   & 0.45         & 2      &  0.38   &  4(c) \\
G296.7$-$0.9  & 11:55:31.0 & $-$63:07:08 & $14\times10$  & 43   & 2.9         & 1      &  2.46   &  5(a) \\
G299.3$-$1.5  & 12:17:55.5 & $-$64:09:53 & $36\times29$  & 12   & $>0.31$   & 2    &  $>0.04$ & 5(b) \\
G308.4$-$1.4  & 13:41:32.9 & $-$63:44:41 & $14\times6$    & 37   & 0.39         & 1.5      & 0.55   &  5(c)\\
G310.7$-$5.4  & 14:12:18.1 & $-$67:05:03 & $31\times29$  & 6     & $>0.83$  & 1   &  $>0.11$  &  6(a)  \\
G310.9$-$0.3  & 13:59:55.5 & $-$62:03:47 & $17\times11$  & 38   & 0.64         & 3     & 0.40  &  6(b) \\
G321.3$-$3.9 & 15:32:13.8 & $-$60:51:54  & $109\times64$ & 10  & $ >0.37$ & 1  &  $>0.01$  &  6(c) \\
G322.7+0.1    & 15:24:01.9 & $-$56:49:12 & $13\times11$  & 7     & 0.29         & 1.5      & 0.24    &  7(a) \\
G322.9$-$0.0  & 15:25:40.4 & $-$56:46:45 & $12\times11$  & 7     & 0.17         & 1.5      & 0.15   &  7(b) \\
G323.7$-$1.0  &  15:34:30.1 & $-$57:12:03  & $51\times38$   & 6   & $>0.61$   & 1  & $>0.04$   & 7(c)  \\
G324.1+0.0   & 15:32:41.2 & $-$56:03:14 & $11\times7$    & 36   & 0.30         & 2      & 0.46      &  8(a) \\
G325.0$-$0.3  & 15:39:20.7 & $-$55:49:28 & $6\times4$      &  37  & 0.69         & 2    & 3.40    &  8(b)  \\
G330.7+0.1    & 16:07:28.4 & $-$51:53:08 & $10\times8$     &  32   &  -- & 2  &  --    &  8(c)    \\
G334.0$-$0.8  & 16:26:36.0 & $-$50:16:19 & $14\times11$   & 27  & 0.23   & 2     & 0.18   &  9(a)     \\
G336.7$-$0.3  & 16:35:25.4 & $-$47:57:01 & $5\times3$       &  72  & 0.66   & $>3$  & 5.20   &  9(b)     \\
G336.9$-$0.5 & 16:37:12.2  &  $-$47:57:53 & $15\times12$  & 26  &   -- & 3 &  -- &  9(c)  \\
G345.1$-$0.2  & 17:05:14.1  & $-$41:26:00 & $6\times5$       & 172  &  2.1        & 3     & 0.59  &  10(a) \\
G345.2+0.2  & 17:03:44.3 & $-$41:05:35 & $11\times11$    & 19   & 0.28    & 2     &  0.27  &  10(b)   \\
G346.2$-$1.0  & 17:12:10.8  & $-$40:59:06 & $7\times7$       & 14    & 0.55   & 2     & 1.33  &  10(c)  \\
G348.9+1.1  & 17:11:28.7 & $-$37:34:42 & $12\times11$   & 17    &  2.2         & 3    &2.00 & 11(a)    \\
G354.1+0.3  & 17:29:38.4  & $-$33:47:02 & $11\times11$   & 52    & 0.11  & $>3$  & 0.11  &  11(b)   \\

\hline\hline 
\end{tabular}
\end{center}
\label{tab2}

\end{table*}

The {\it Midcourse Space Experiment} ({\it MSX}; Egan \& Price 1996, Price et al. 2001) survey at 8$\mu$m is well-matched in angular resolution and sensitivity to the MGPS-2 and robust comparison of currently confirmed SNRs (Green 2009, 2014) finds that no SNRs contradict the anticorrelation (Cohen \& Green 2001). A blind comparison between MGPS-2 and the {\it MSX} 8$\mu$m images recovered all the known SNRs to date, confirming the validity of the process. The {\it MSX} survey extends in latitude $|b|\leq5^{\circ}$ so that the outer strips of $5^{\circ}\leq|b|\leq10^{\circ}$ of the MGPS-2 survey have not been searched using our selection criteria. For example, the recent discovery of SNR G351.0$-$5.4 (De Gasperin et al. 2014) is located outside our blind search region. Table 2 lists the 23 candidate SNRs identified by our procedure. 

Some of the sources have been previously proposed as potential candidates but have not yet been confirmed and were not included in the SNR catalogue of Green (2009). This is because of limited data quality or from the lack of any other available radio or infared data of sufficient sensitivity or angular resolution. The candidates from Duncan et al. (1997) have an angular resolution of $10^{\prime}$ and a sensitivity of 17 mJy. The present data are more than a factor of 10 improved in both parameters. The possible candidates from Whiteoak \& Green (1996), listed in MSC.C, are included because no images were provided and the MOST telescope artefacts have been reduced with the new widefield mode. 

The search in this paper was conducted as a blind survey. However, the SNR catalogue of Green (2014) has in its documentation about 20 sources in our search area listed as possible candidates, proposed on the basis of their radio emission. For completeness, these sources were also reviewed using the same selection criteria listed above. Six of the sources were smaller than our search diameter; of the remaining sources, all but one are located in areas of highly structured MIR emission, making a clear identification impossible without more detailed followup observations. The one source that looks promising is G358.7+0.7, first proposed by  Gray (1994). This source is weak (peak flux 15 mJy~beam$^{-1}$) and about $17^{\prime}$ in diameter. The source is located close to the Galactic Centre and the MOST image is corrupted by artefacts produced by nearby strong emission. It has not been included in this paper and followup observations should clarify the identification.   

Two sources from our search region are now included in the updated SNR catalogue of Green (2014). G296.7$-$0.9 (Robbins et al. 2011) has been published with new multiwavelength data. De Horta et al. (2013) have analysed archival data for G308.4$-$1.4 to propose only the western ellipse of the source is an SNR. Both are included to record the  completeness of our search results. None of the sources listed in Table 2 are in the SNR catalogue of Green (2009). It is expected the candidate SNRs will be a useful guide for further study with the low frequency Murchison Widefield Array (MWA; Tingay et al. 2013, Bowman et al. 2013), where continuum confusion limits of more than 10 mJy might otherwise restrict identification of weaker sources.

For Table 2, column 1 gives the Galactic name (and coordinates), columns 2 and 3 the J(2000) Right Ascension and Declination, column 4 the source size given as half-power major and minor axes in arcmin, columns 5 -- 7 give the peak flux in mJy beam$^{-1}$, the total integrated flux density in Jy and the 1$\sigma$ noise in the area surrounding the candidate source, in mJy beam$^{-1}$, respectively. Column 8 is an estimate of the surface brightness of the source in units of $10^{-5}$ Jy sr$^{-1}$, equivalent to $10^{-21}$ W m$^{-2}$Hz$^{-1}$sr$^{-1}$. It is a lower limit for sources $\geq25^{\prime}$ in diameter, which are not fully measured by MOST. Column 9 gives the postage stamp figure number. 
The positional uncertainty of the sources is a few arcsec, largely determined by the lack of precision in determining the centroid of the sources with their irregular morphology. The nominal error in the flux densities is between 5 and 10\%. Some of the figures are very weak but still at least $3-5\sigma$ times the rms noise floor.

\begin{figure*}
\begin{center}
%
\includegraphics[width=5cm]{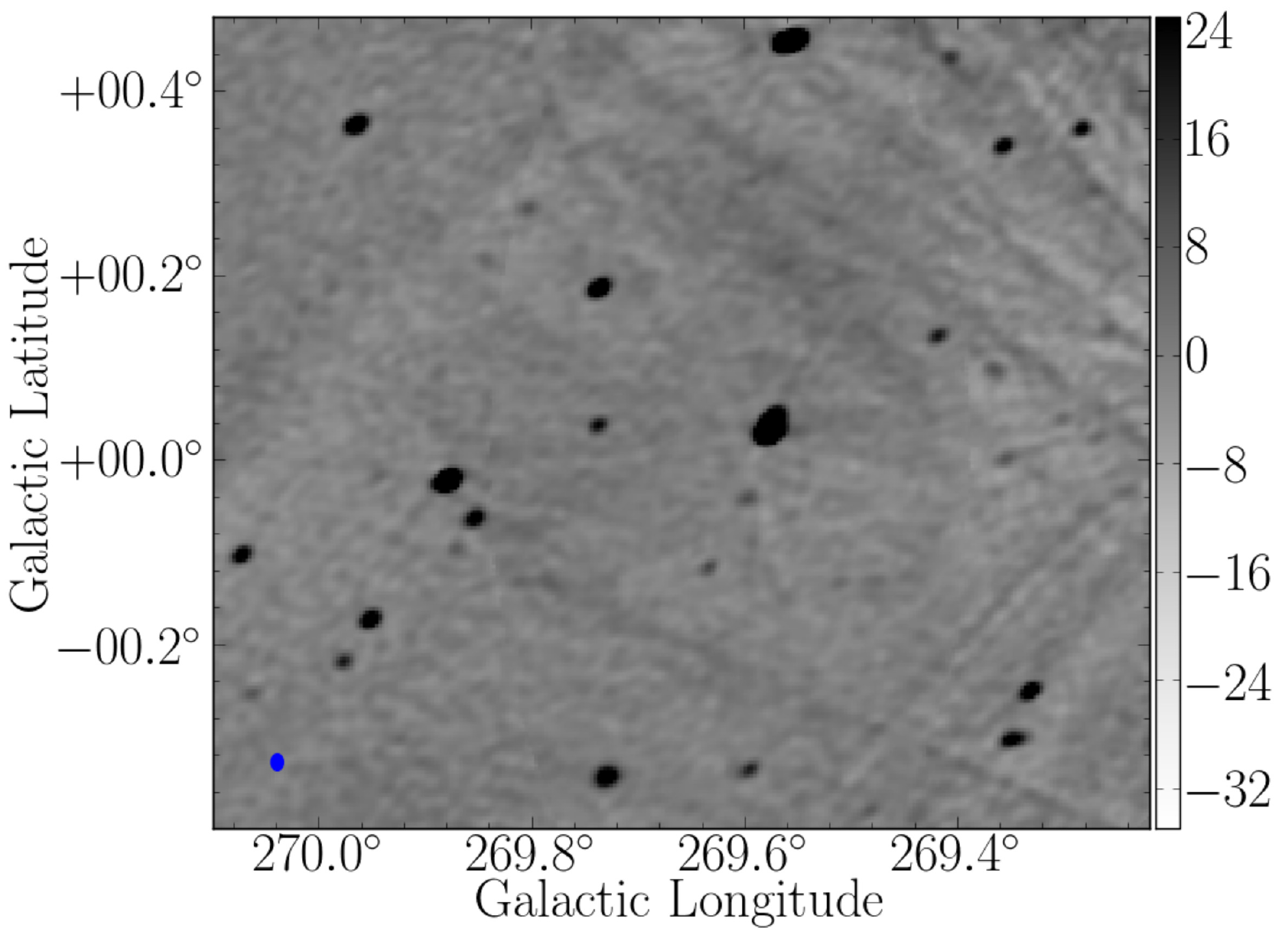}
%
\includegraphics[width=5cm]{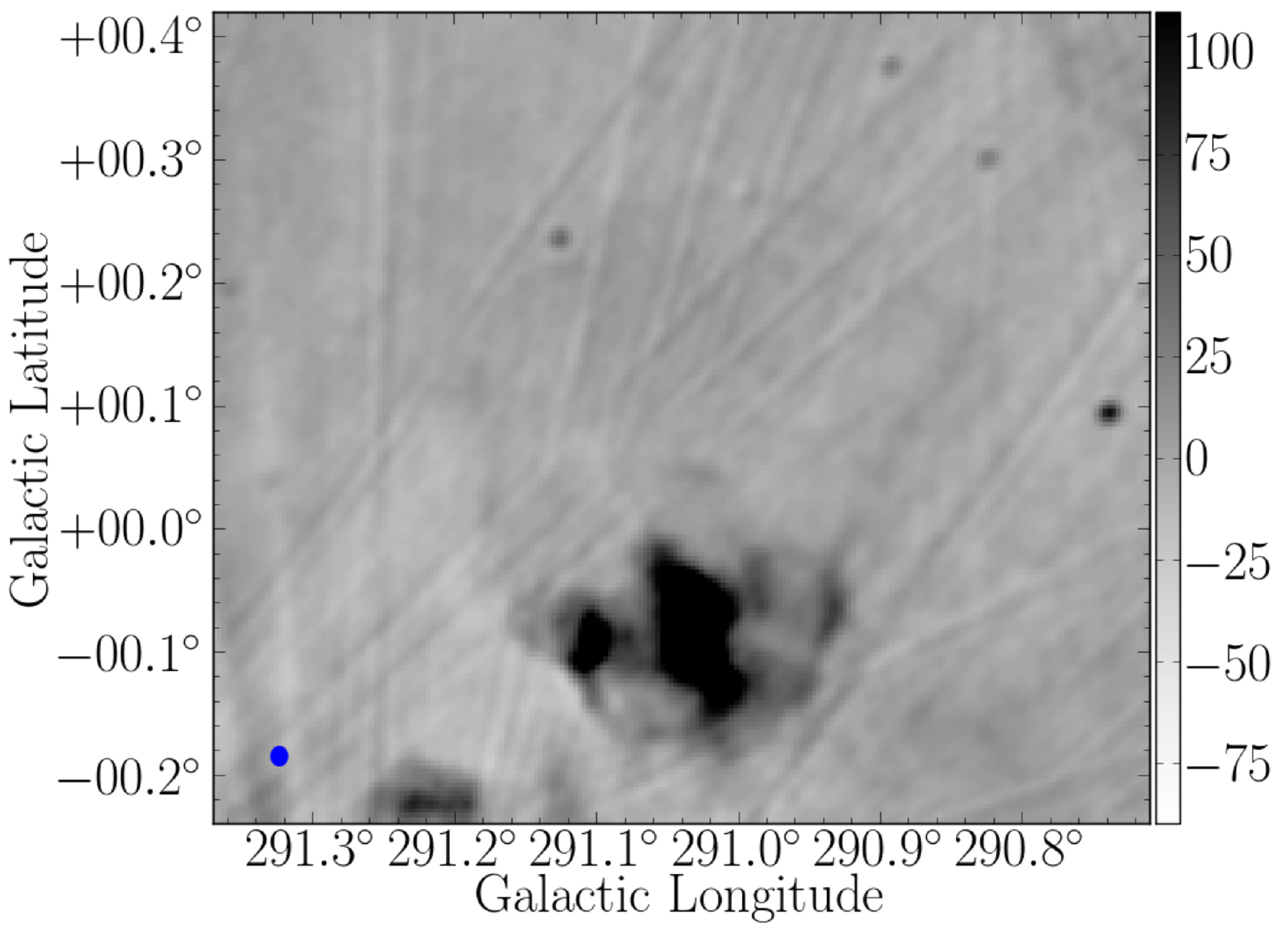}
%
\includegraphics[width=5cm]{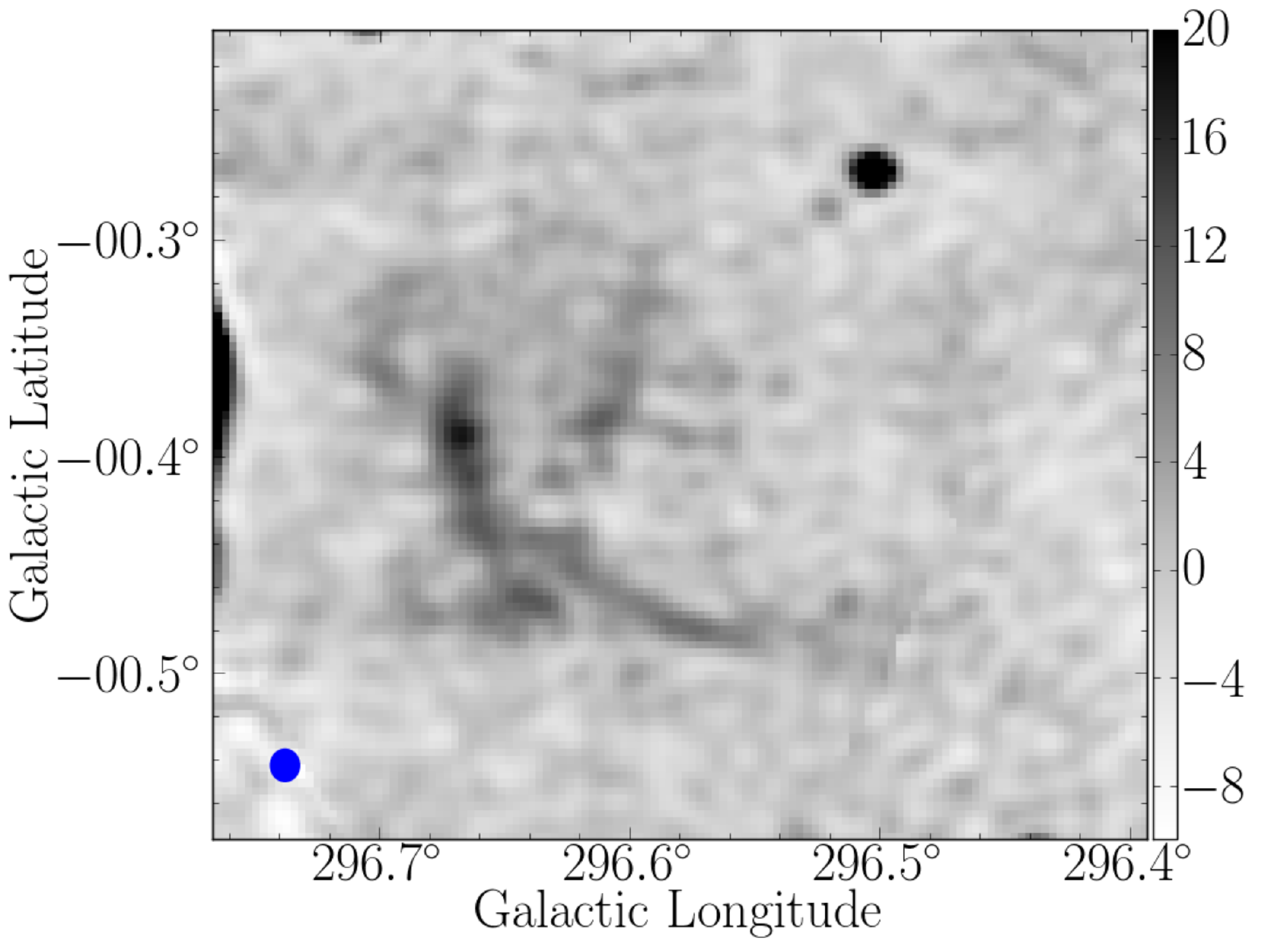}
\caption{ (a) G269.7+0.0 (b) G291.0+0.1 (and G291.0--0.1) (c) G296.6--0.4}\label{snrs-C1,C2,C3}
%
\includegraphics[width=5cm]{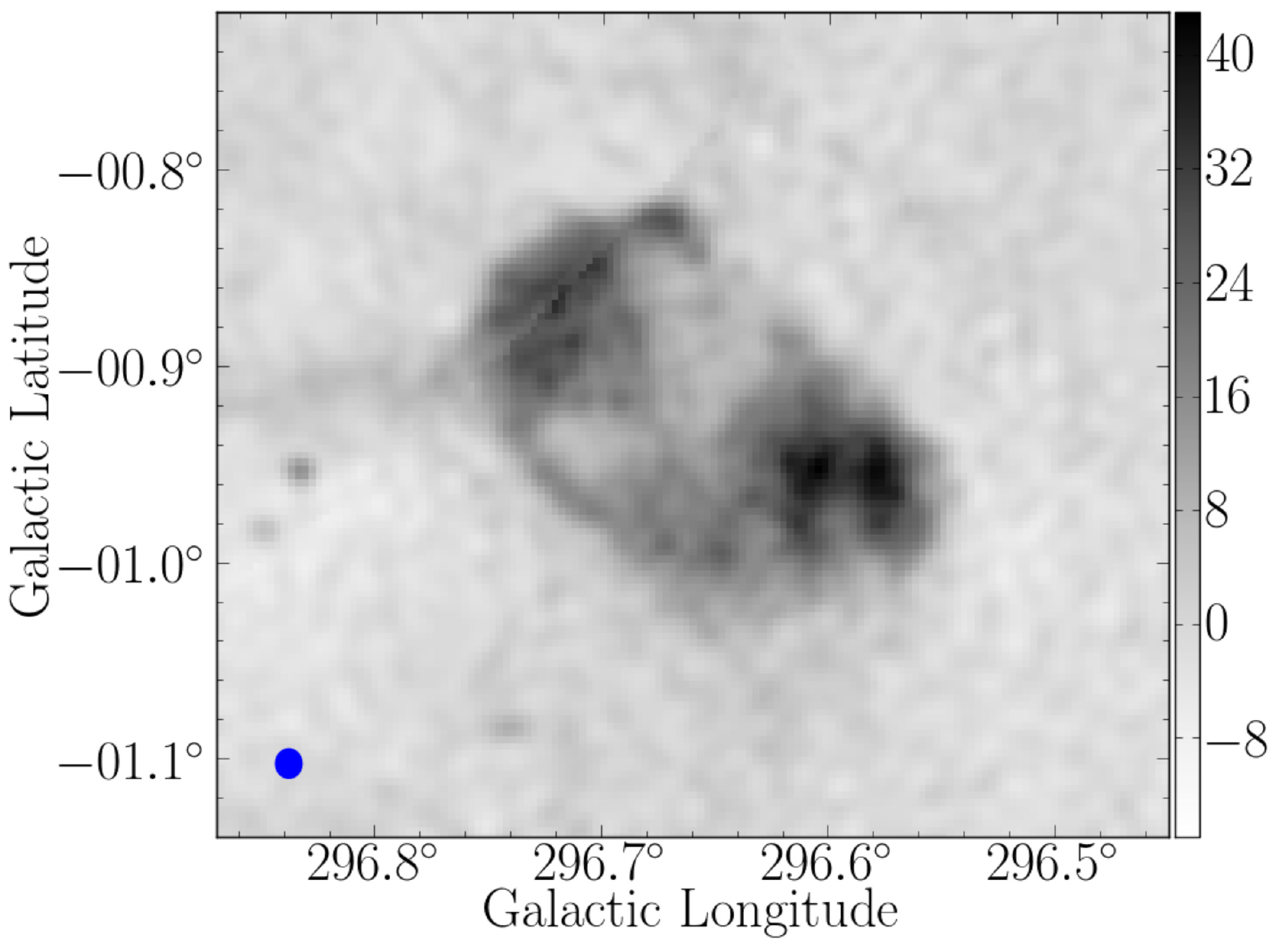}
%
\includegraphics[width=5cm]{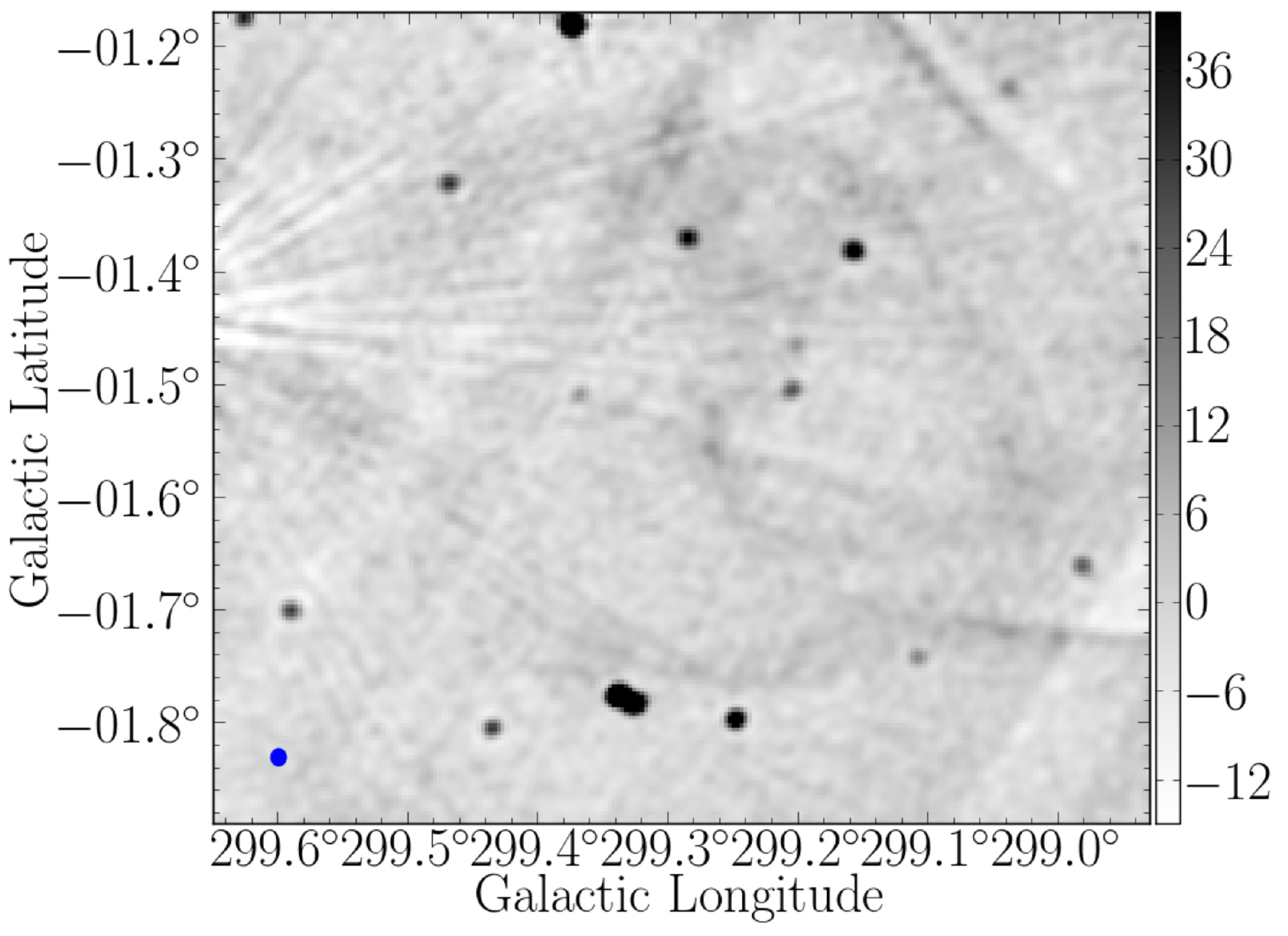}
%
\includegraphics[width=5cm]{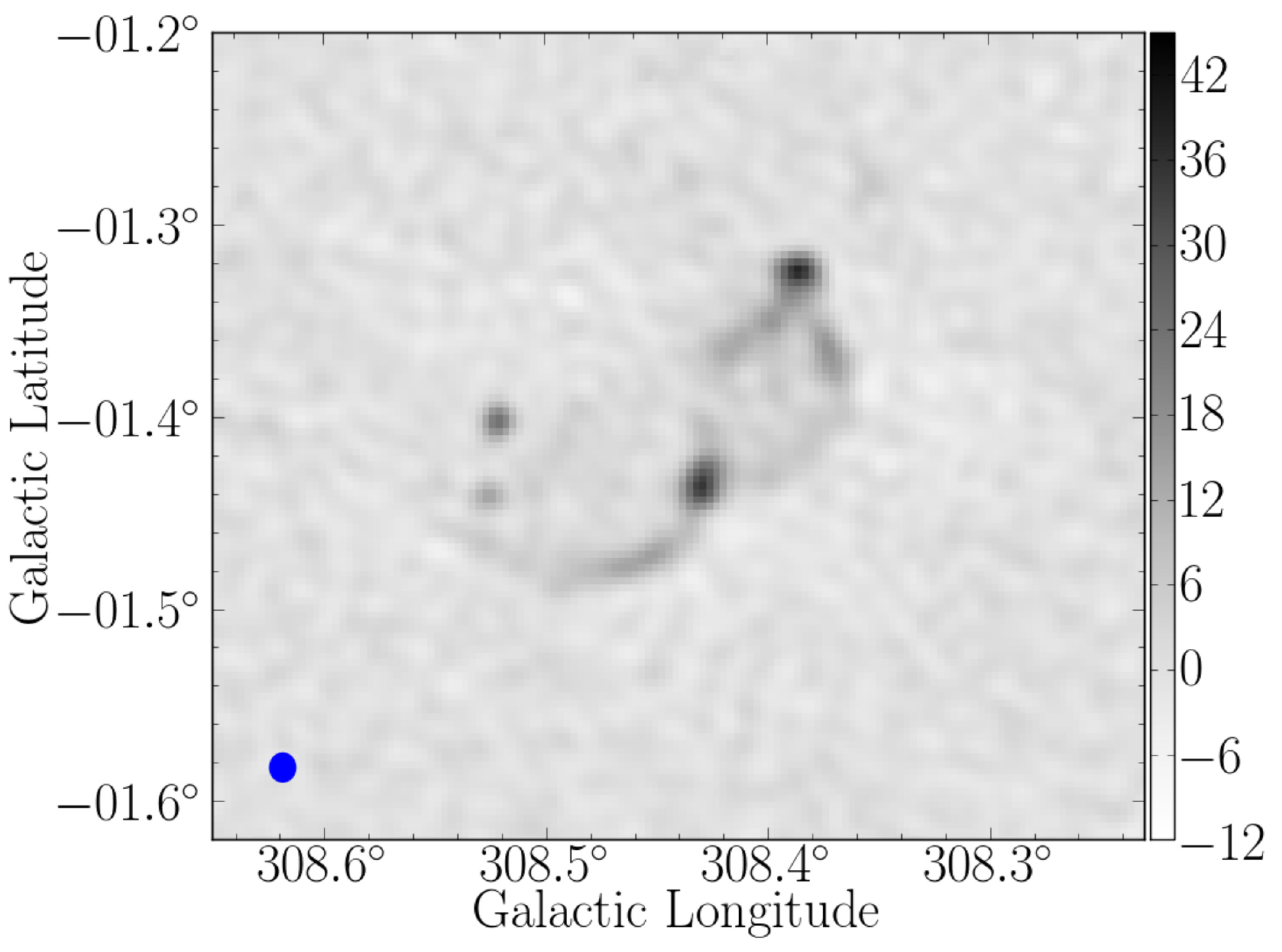}
\caption{ (a) G296.7--0.9 (b) G299.3--1.5 (c) G308.4--1.4}\label{snrs-C4,C5,C6}
%
\includegraphics[width=5cm]{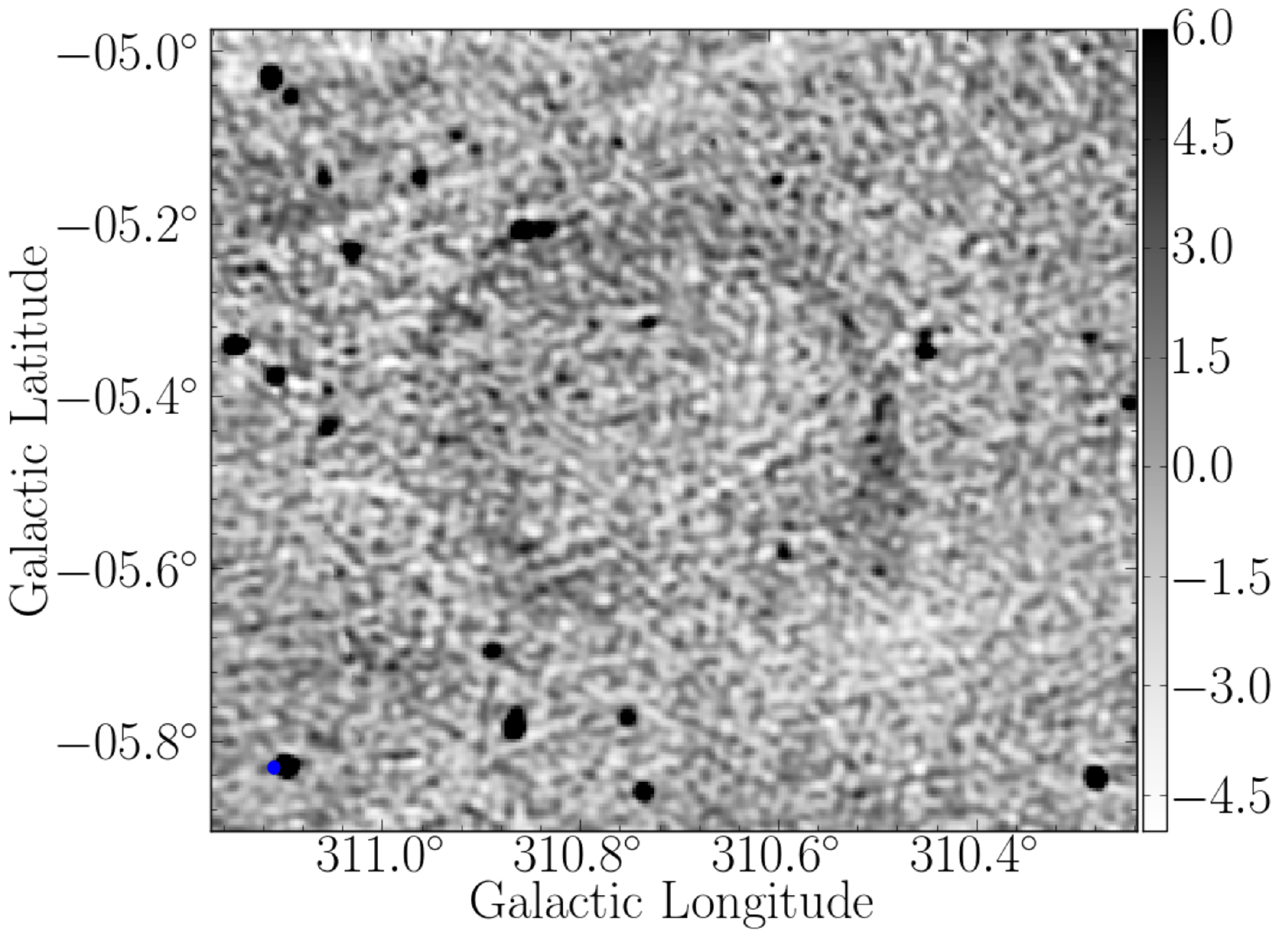}
%
\includegraphics[width=5cm]{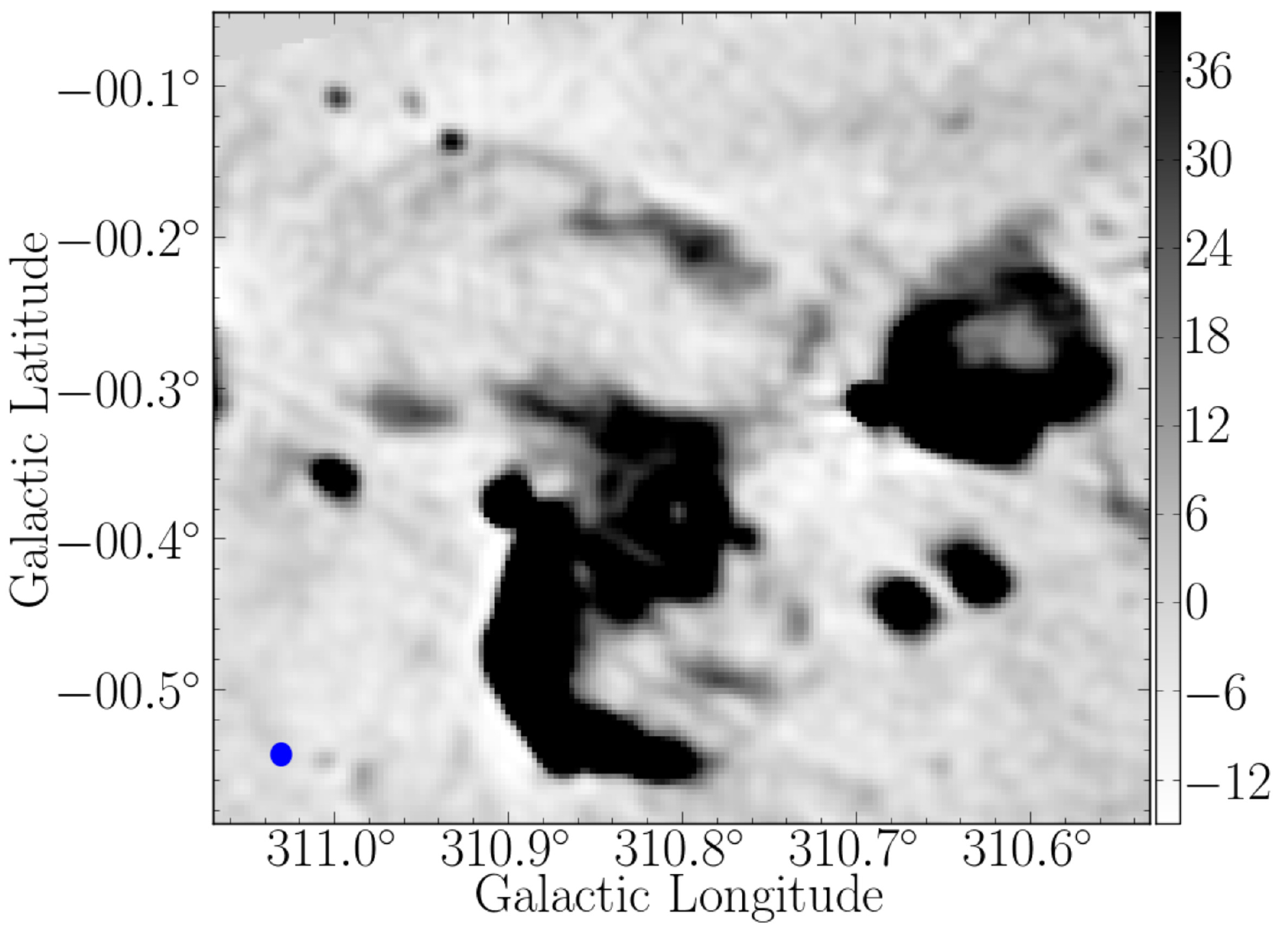}
%
\includegraphics[width=5cm]{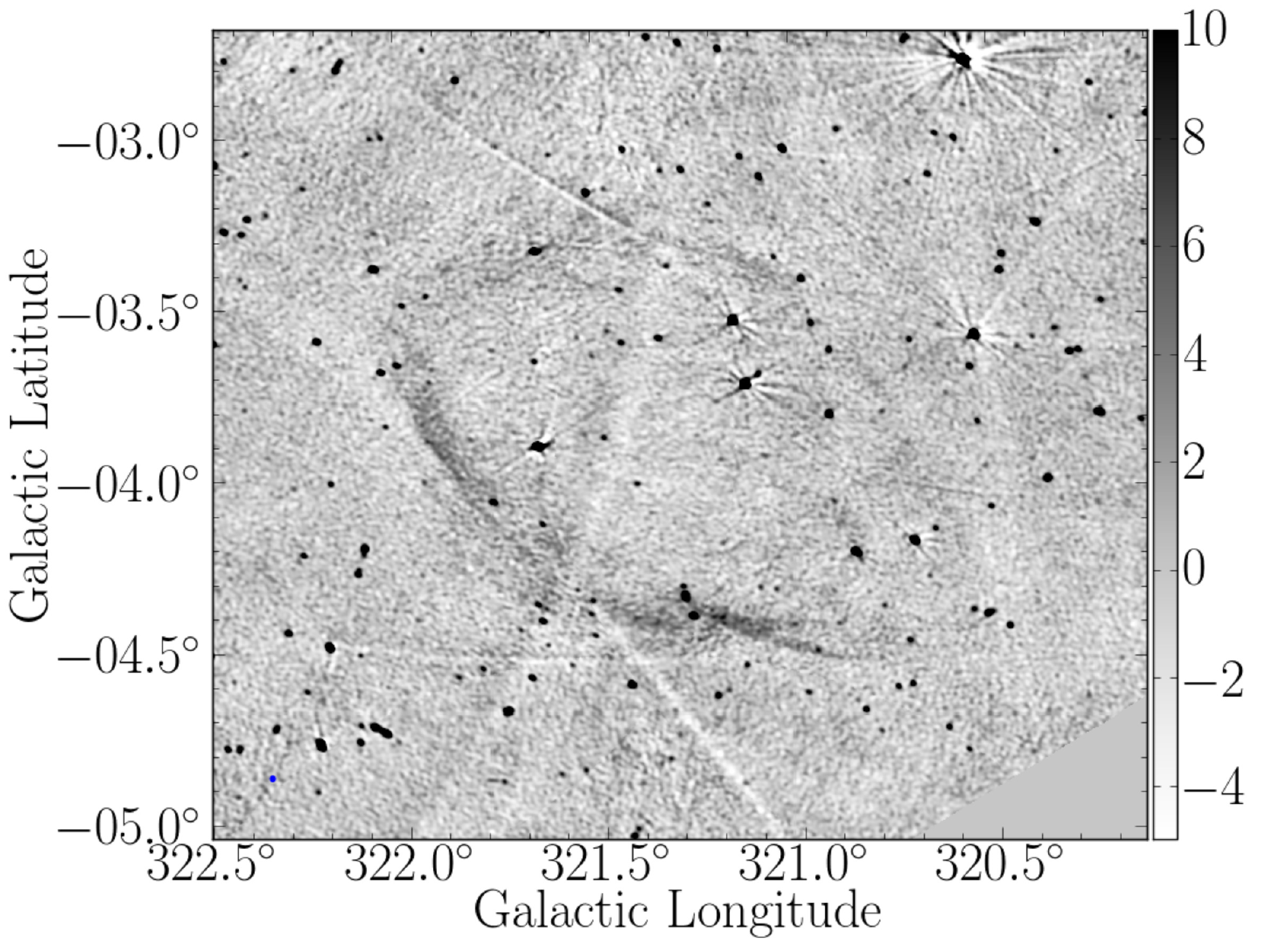}
\caption{ (a) G310.7--5.4 (b) G310.9--0.3 (plus two known SNRs)  (c) G321.3--3.9}\label{snrs-C7,C8,C9}
%
\includegraphics[width=5cm]{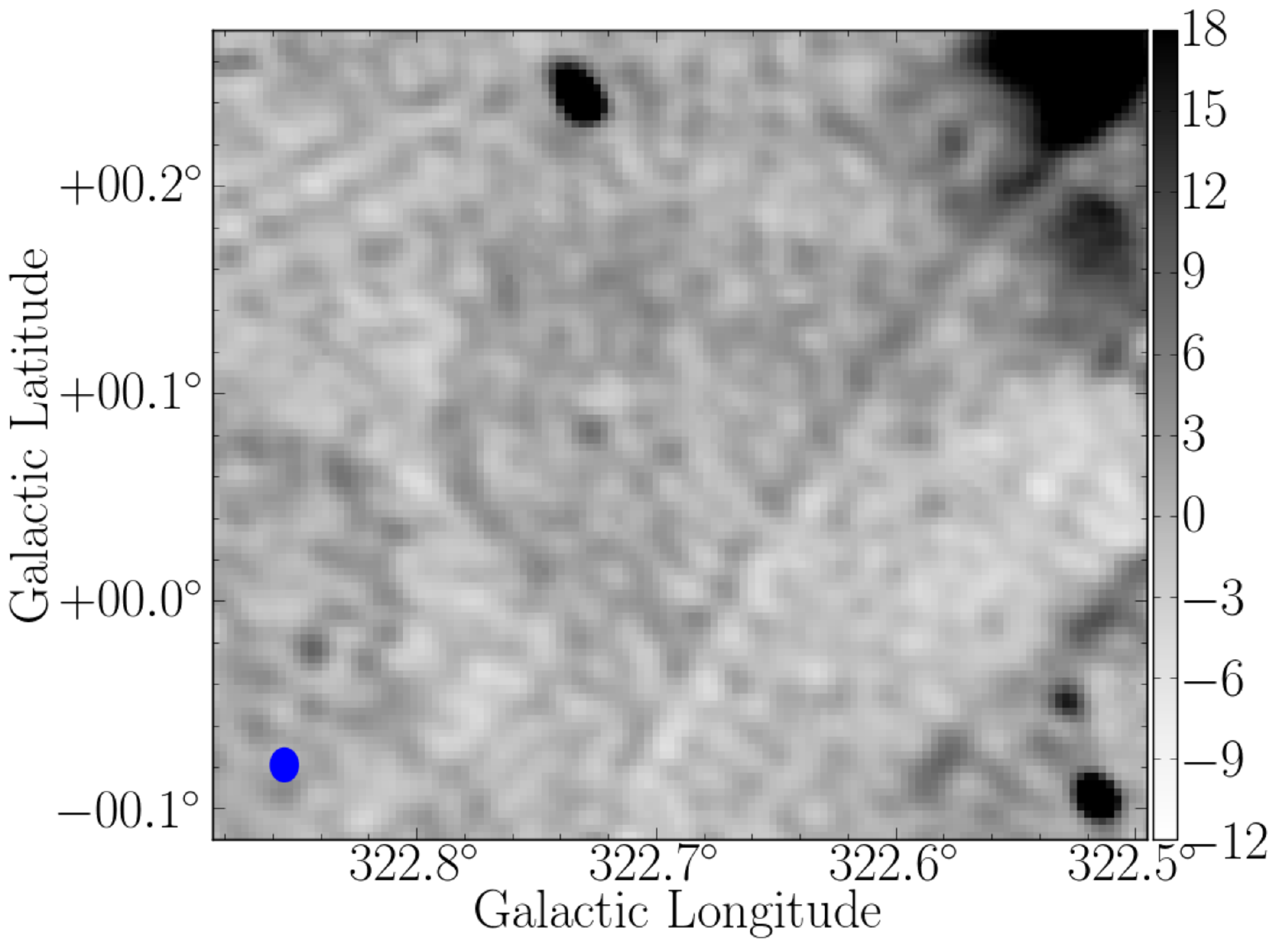}
%
\includegraphics[width=5cm]{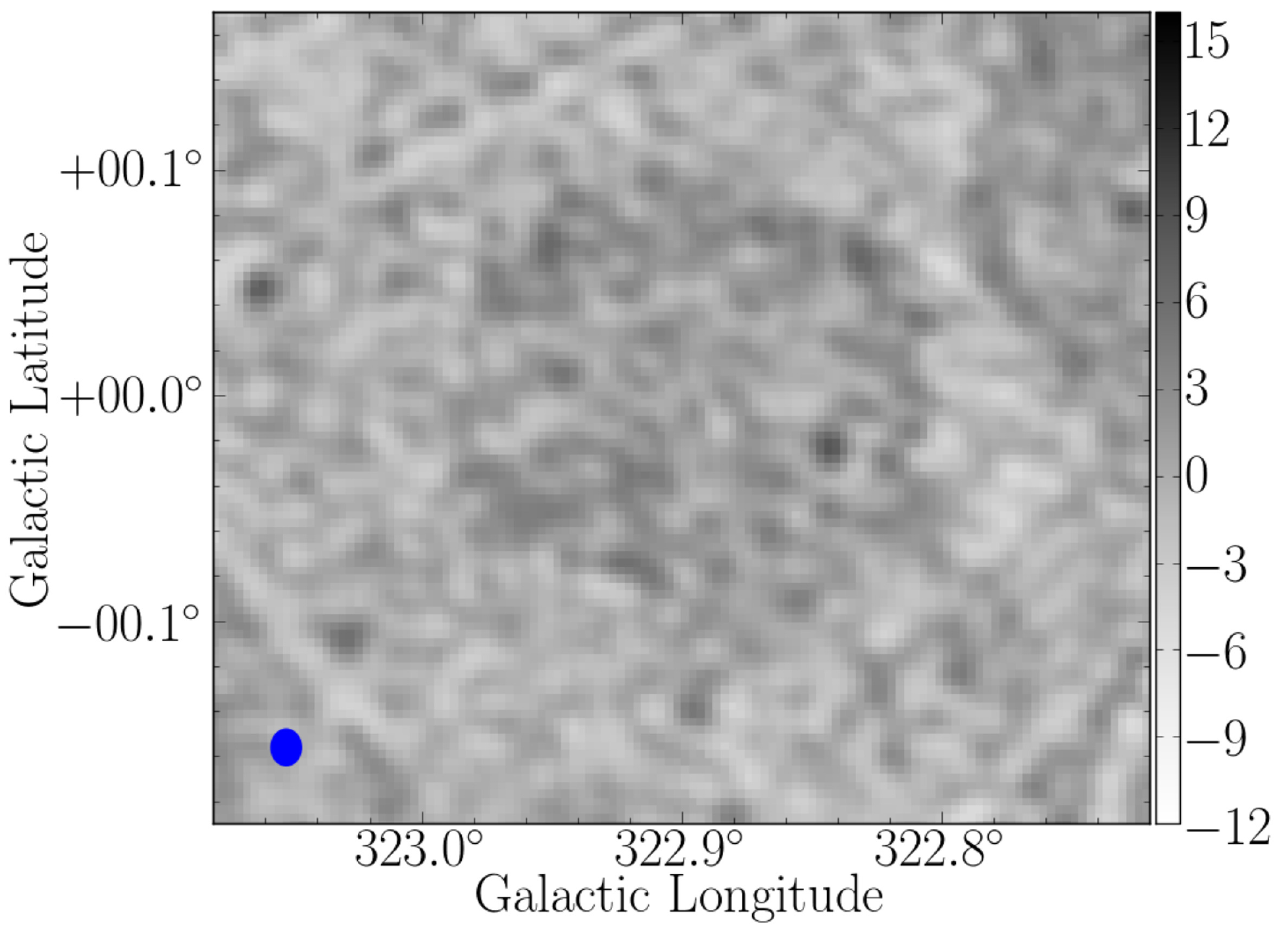}
%
\includegraphics[width=5cm]{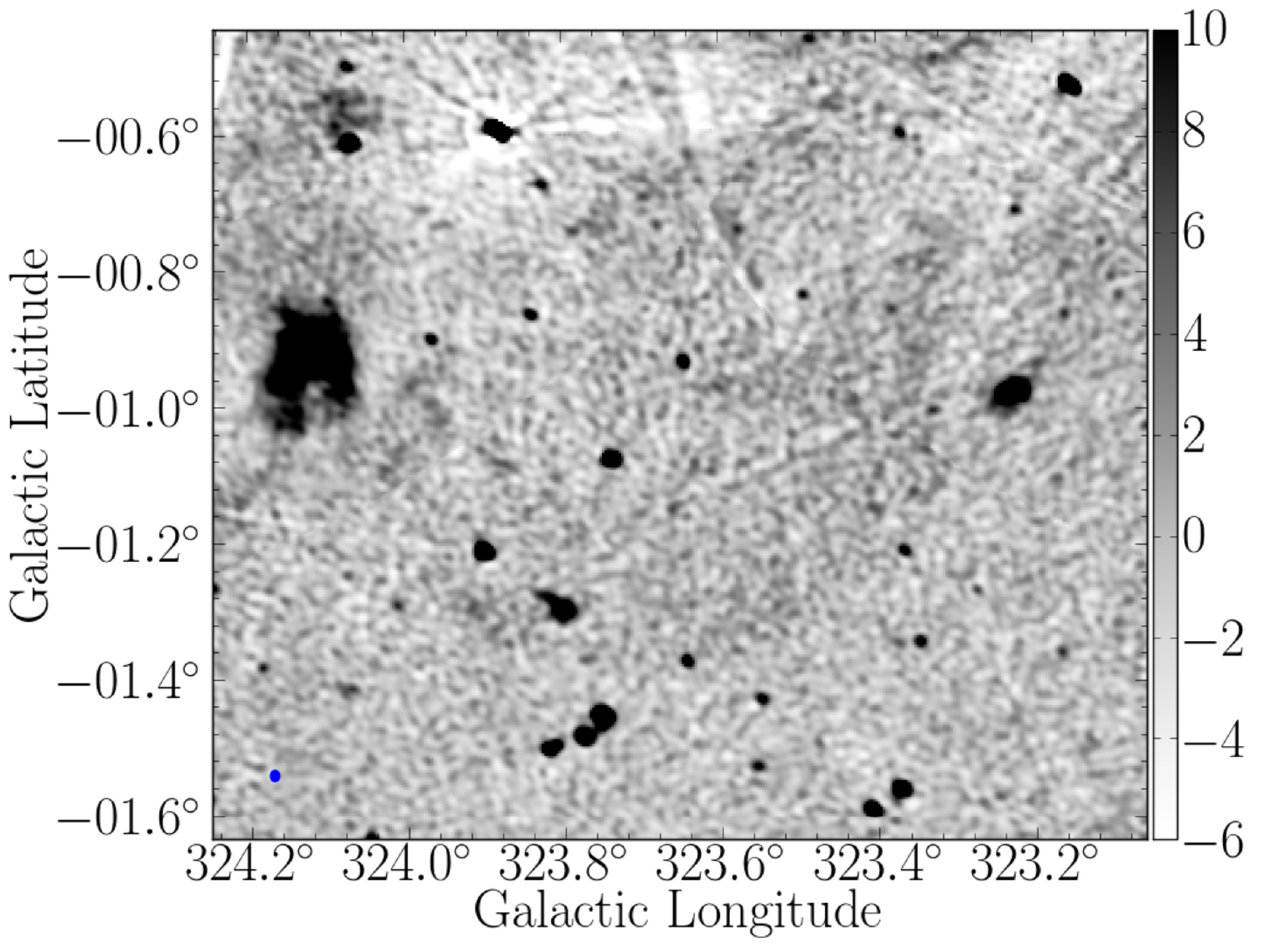}
\caption{ (a) G322.7+0.1 (b) G322.9--0.0 (c) G323.7--1.0 }\label{snrs-C10,C11,C12}
%
\includegraphics[width=5cm]{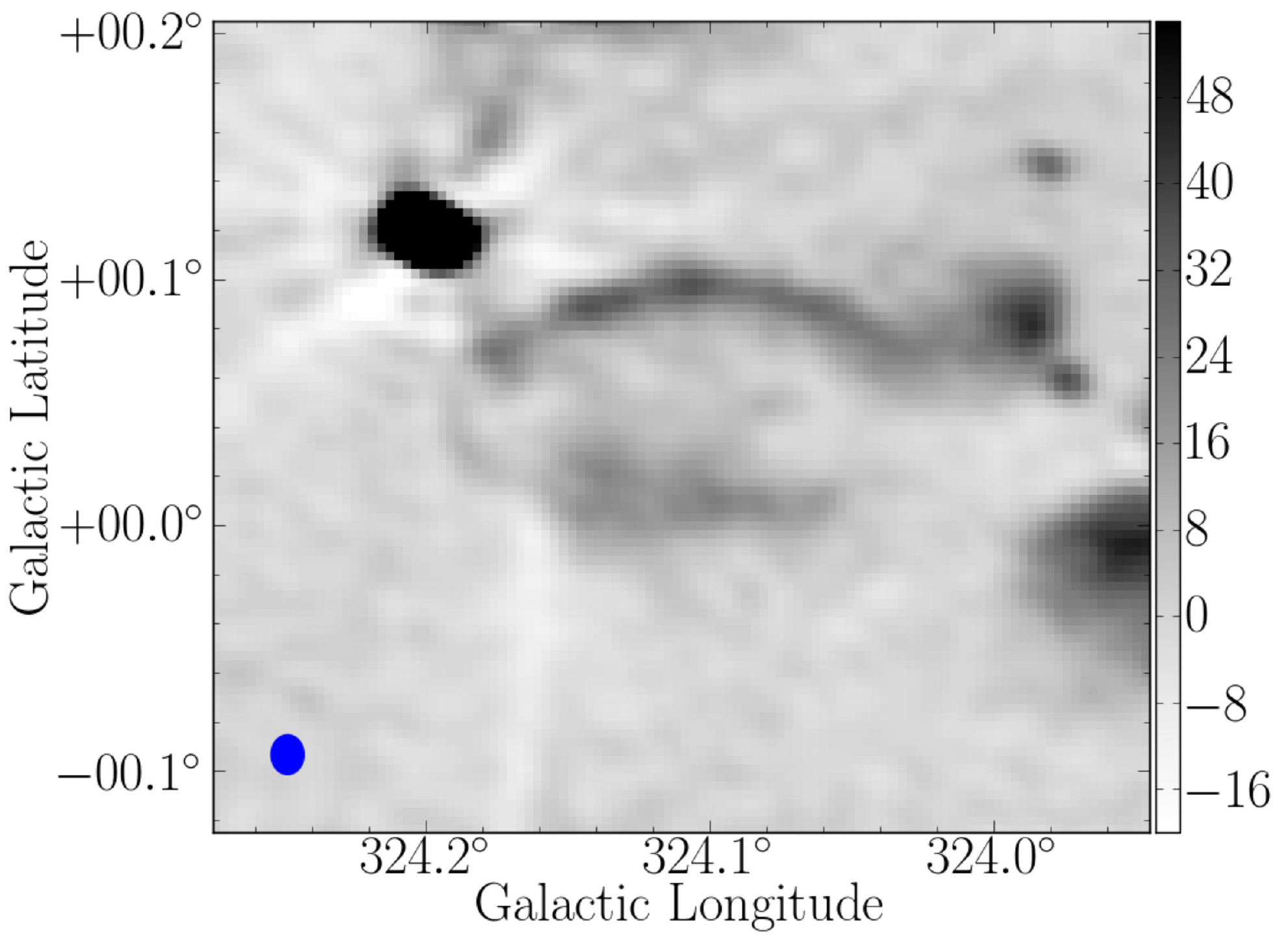}
%
\includegraphics[width=5cm]{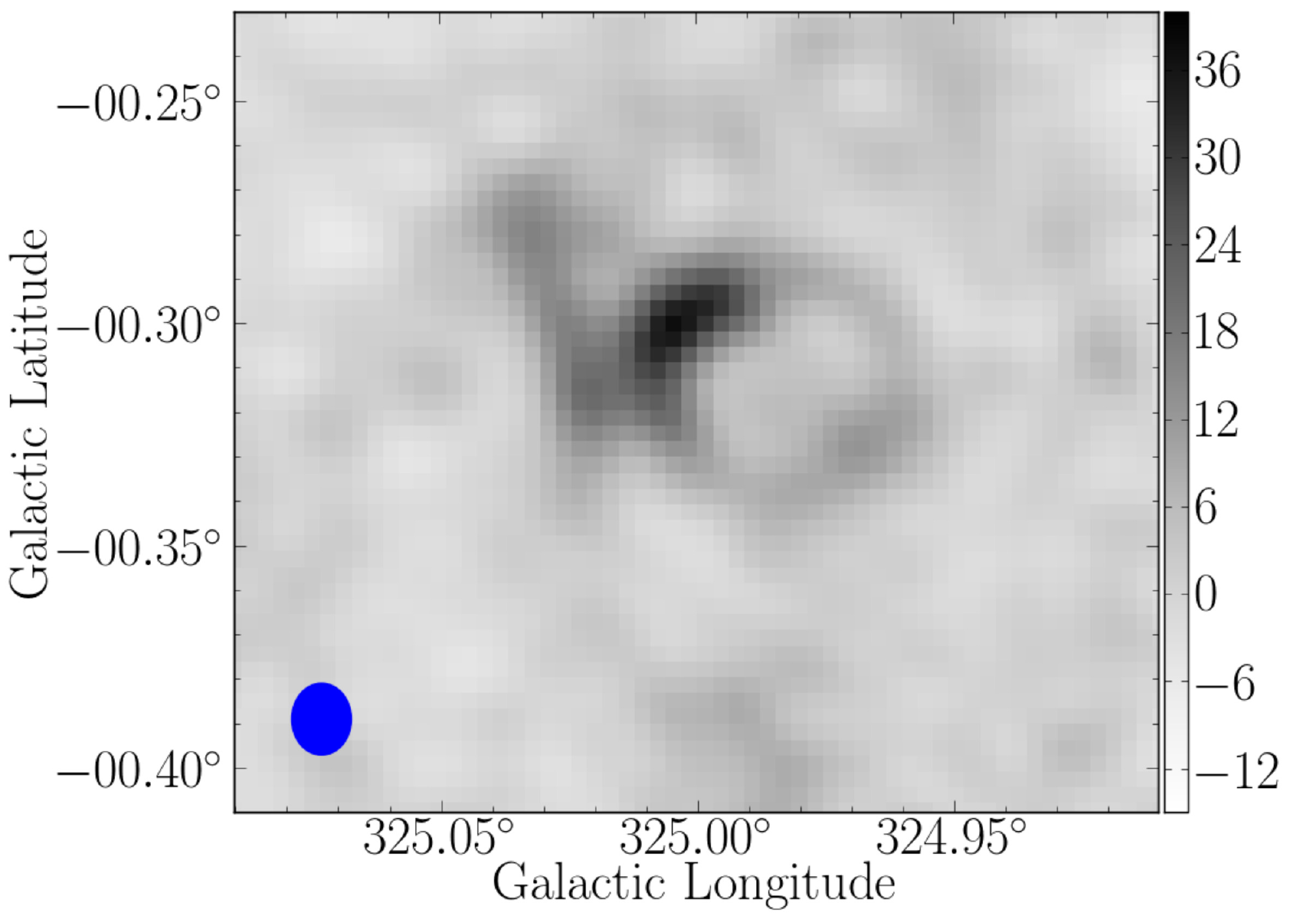}
%
\includegraphics[width=5cm]{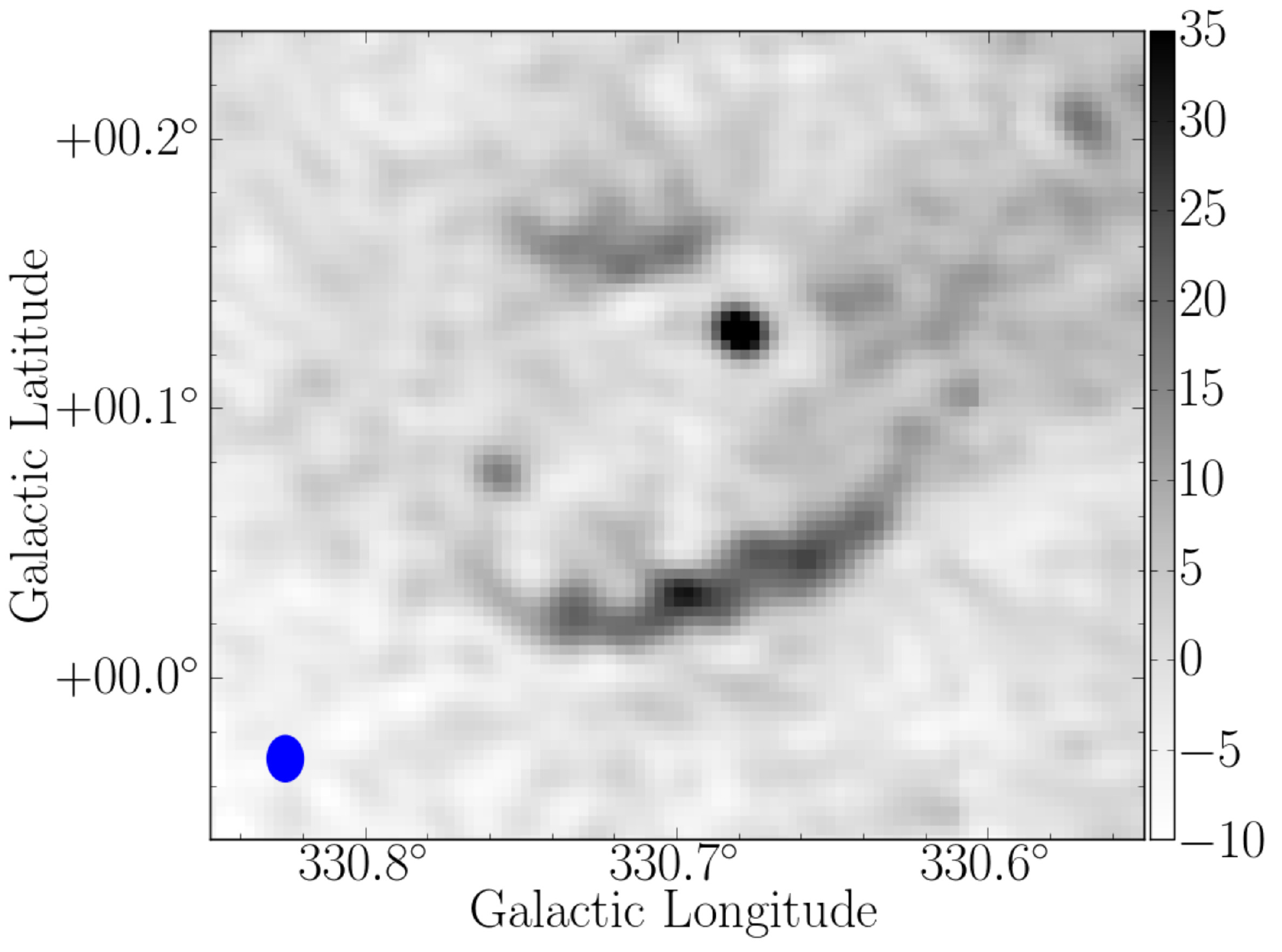}
\caption{ (a) G324.1+0.0 (b) G325.0--0.3 (c) G330.7+0.1 }\label{snrs-C13,C14,C15}
\end{center}
\end{figure*}

\begin{figure*}
\begin{center}
\includegraphics[width=5cm]{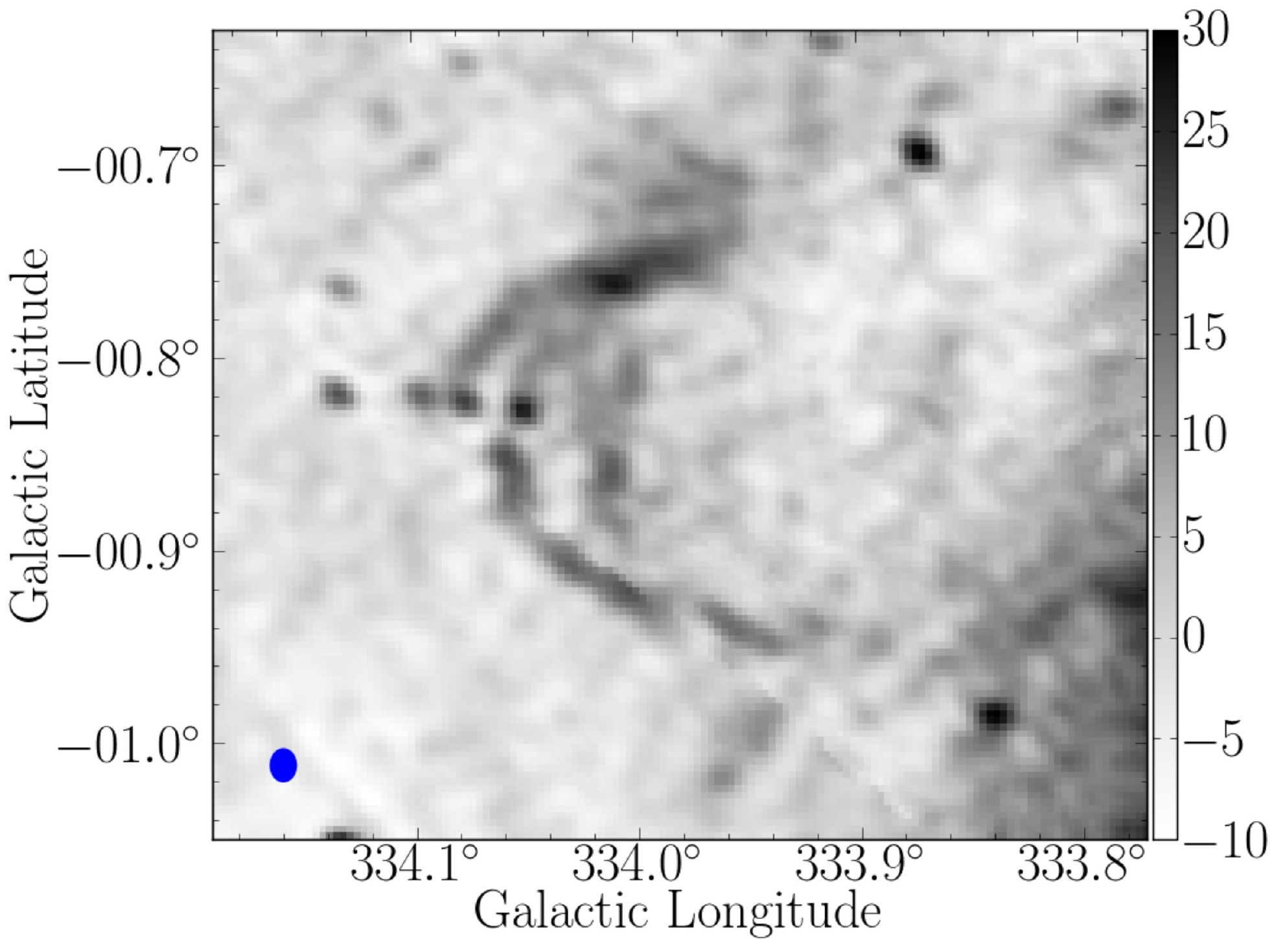}
%
\includegraphics[width=5cm]{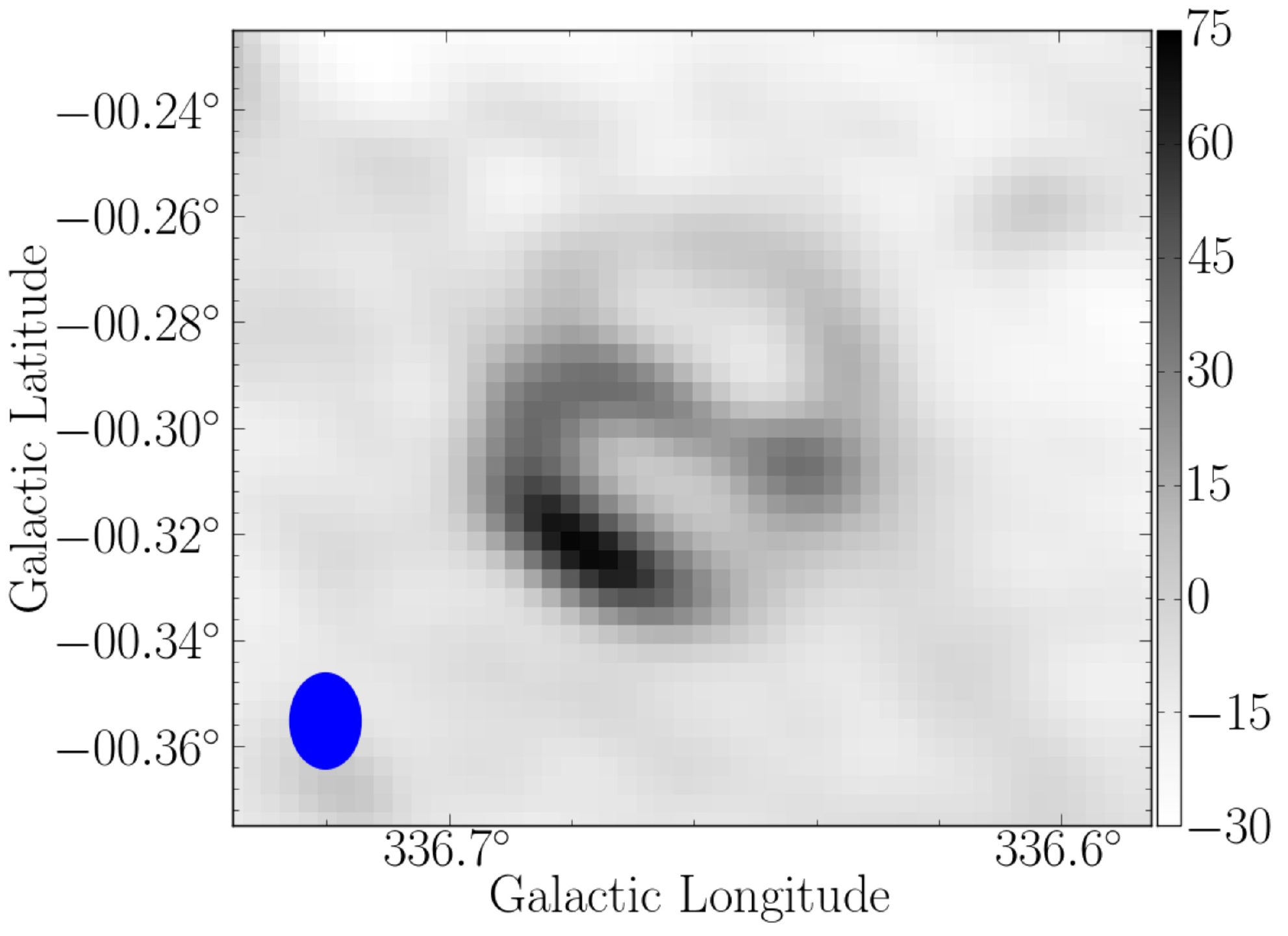}
%
\includegraphics[width=5cm]{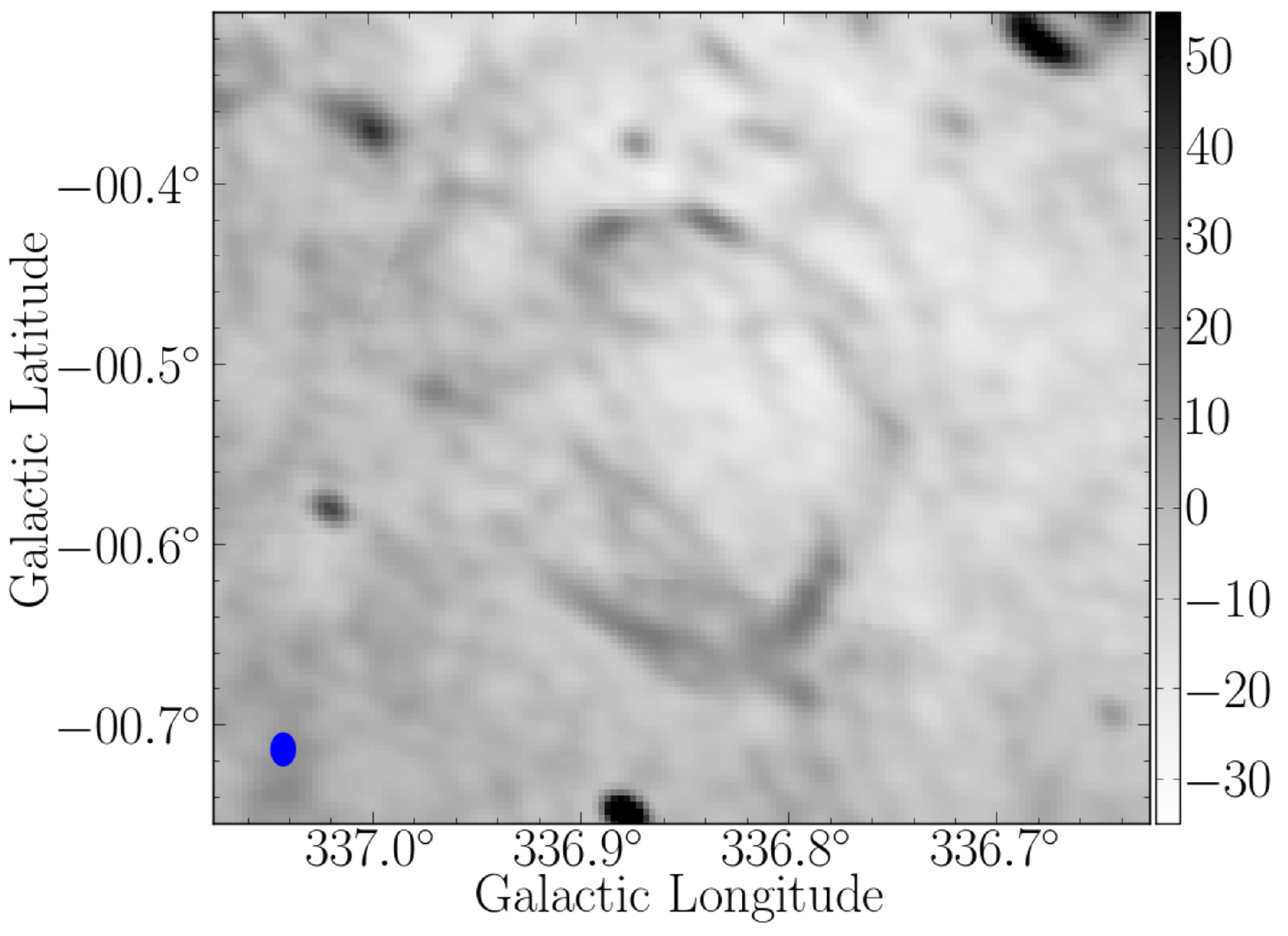}
\caption{ (a) G334.0--0.8 (b) G336.7--0.3 (c) G336.9--0.5 }\label{snrs-C16,C17,C18}
%
\includegraphics[width=5cm]{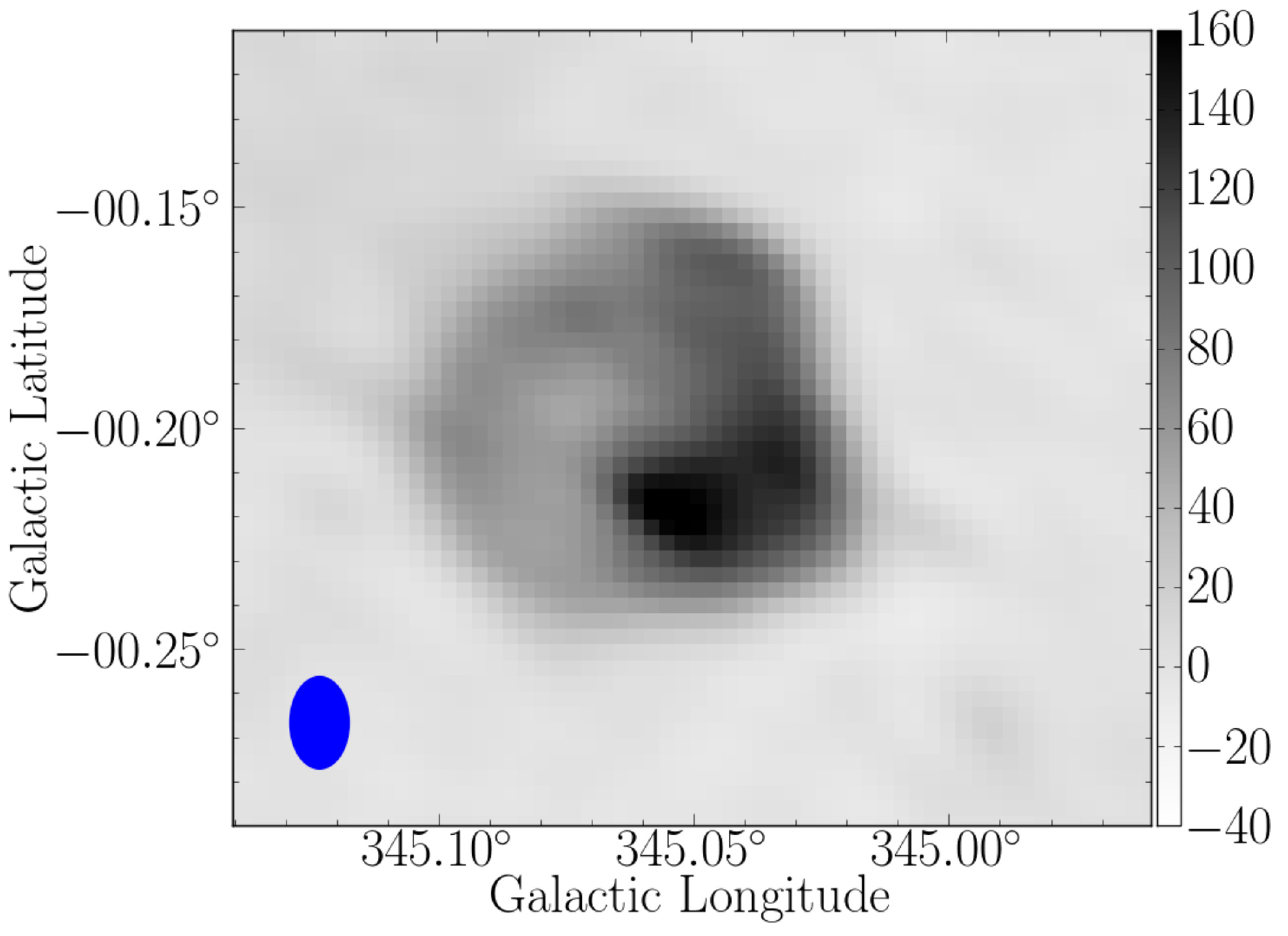}
%
\includegraphics[width=5cm]{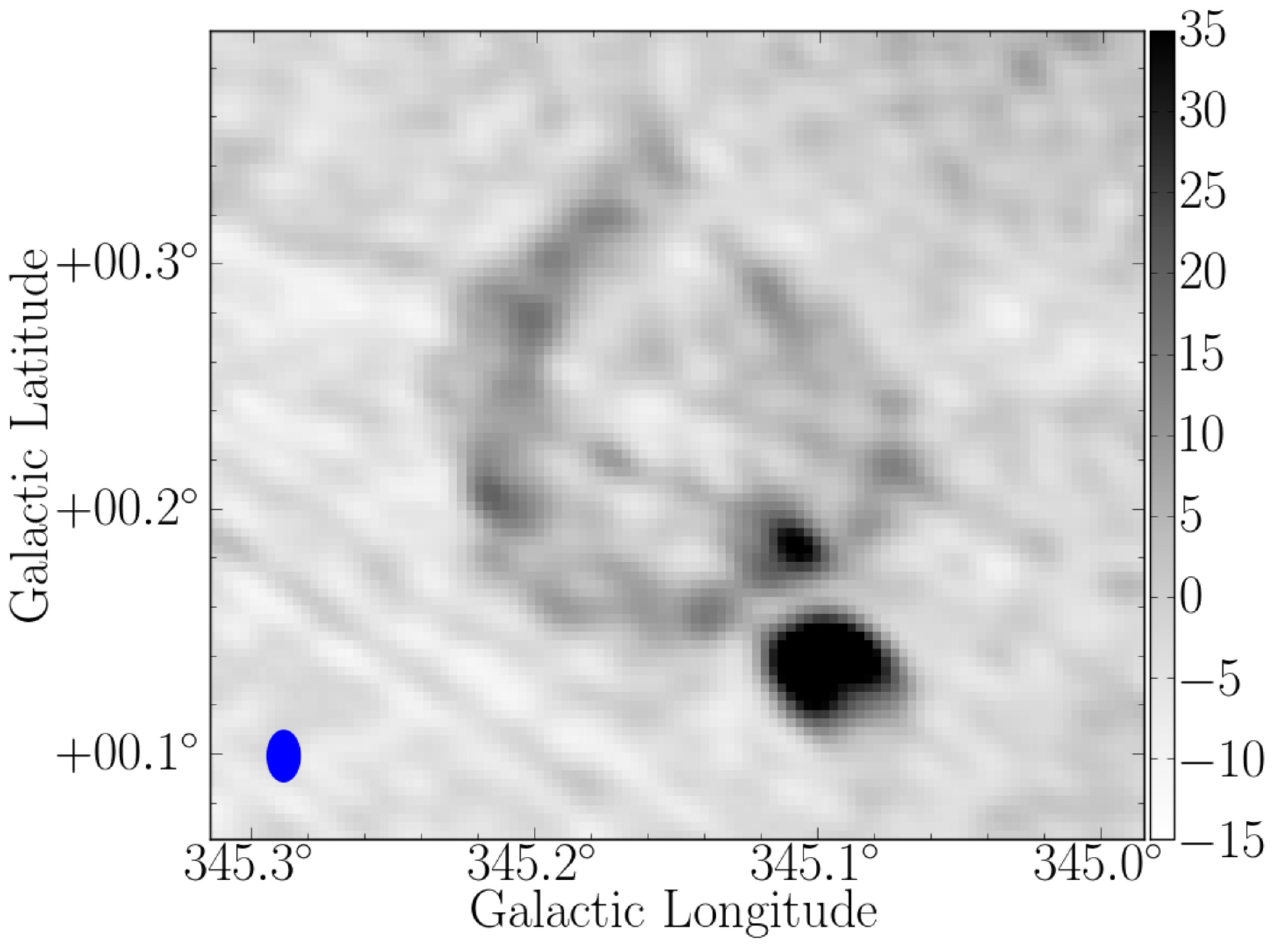}
%
\includegraphics[width=5cm]{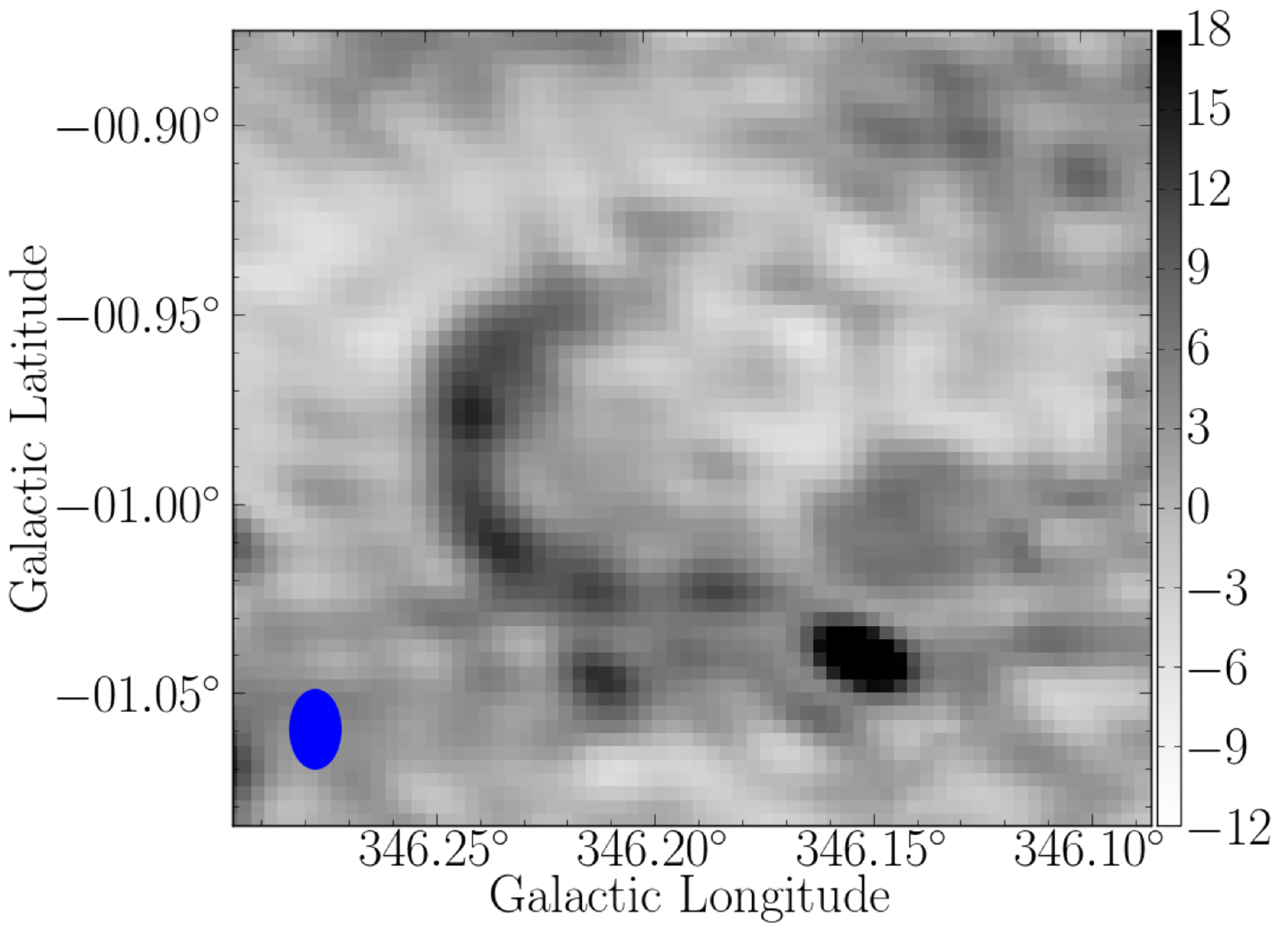}
\caption{ (a) G345.1--0.2 (b) G345.2+0.2 (c) G346.2--1.0 }\label{snrs-C19,C20,C21}
%
\includegraphics[width=5cm]{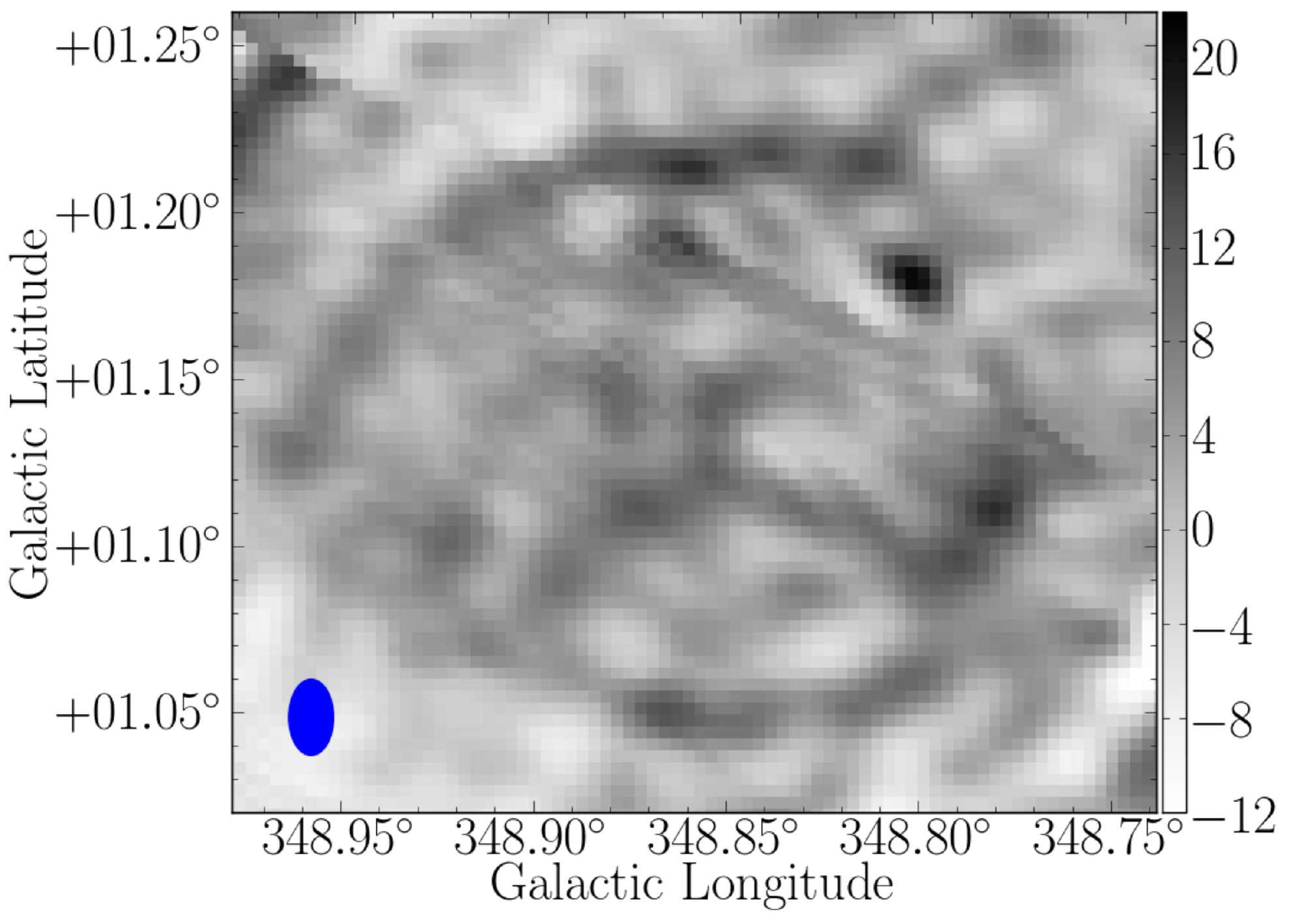}
%
\includegraphics[width=5cm]{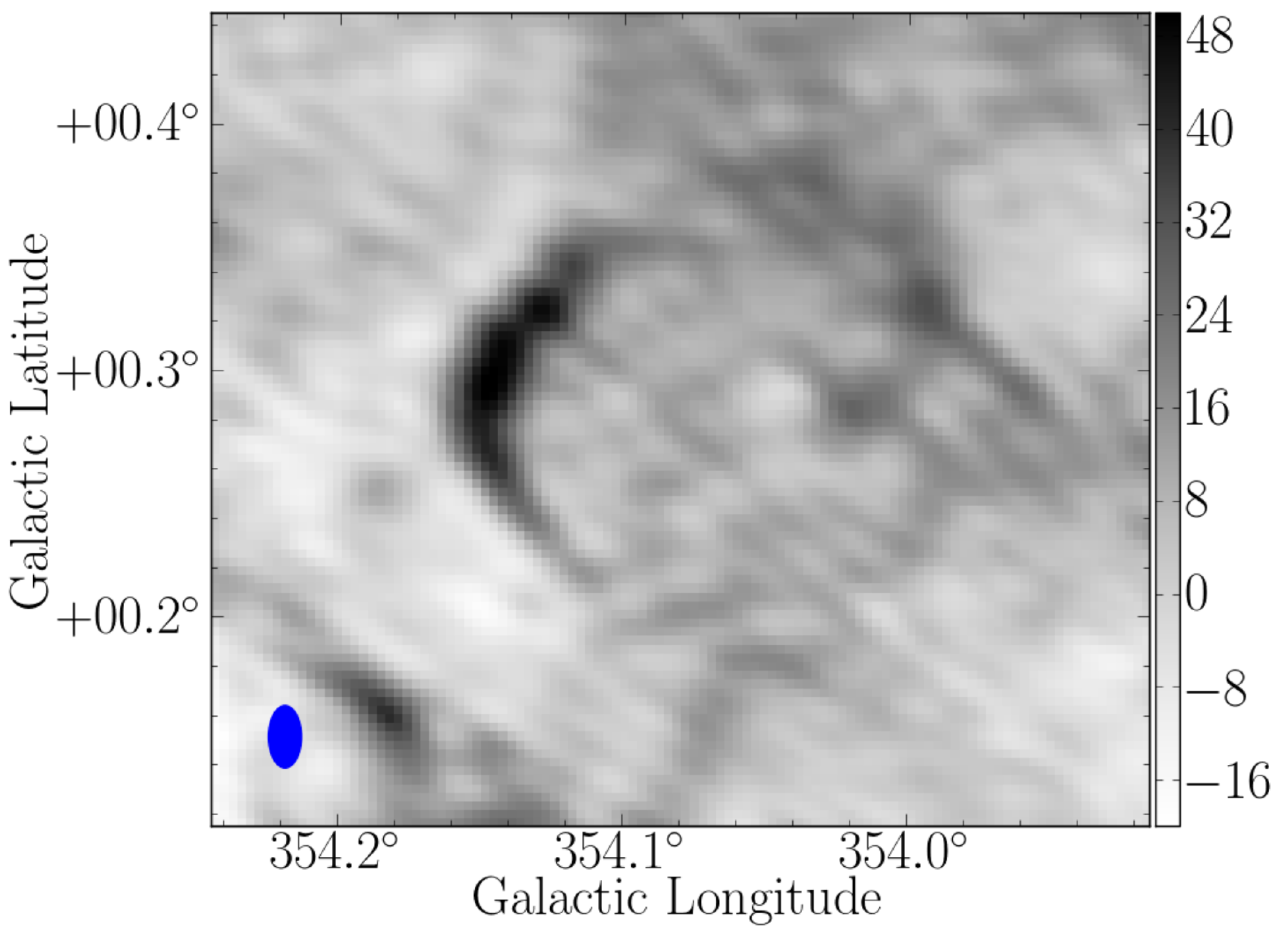}
\caption{ (a) G348.9+1.1 (b) G354.1+0.3 
The colour bar on Figures 4 to 11 gives the greyscale range in mJy beam$^{-1}$ and the telescope beam is shown as a filled blue circle in the lower left corner of each image. }\label{snrs-C22,C23}

\end{center}
\end{figure*}

\subsection{Comments on Individual Candidates}
A brief description of the SNR candidates is given below, noting where the source has been proposed as an SNR candidate in a previous publication.

{\bf G269.7+0.0 [Figure 4(a)]:} Very faint complete circular  shell, affected by imaging artefacts from the massive nearby star-forming region RCW 38. 

{\bf G291.0+0.1 [Figure 4(b)]:} Very faint shell, positioned next to known SNR G291.0-0.1, also shown in the Figure. The morphology of the emission appears to link the sources as part of a single object with a bright southern region, currently identified as a composite SNR.  The northen shell is well aligned with the structures around the composite core. The lack of distance information limits a conclusive interpretation.

{\bf G296.6--0.4 [Figure 4(c)]:} The source is a partial shell of overlapping arcs, separated by about $6^{\prime}$ from the known SNR G296.8--0.3, positioned at slightly higher longitude. Both sources appear to have similar morphology and orientation but there is no evidence of association.

{\bf G296.7--0.9 [Figure 5(a)]:} This source has been published by Robbins et al. (2011) and its identification confirmed. 

{\bf G299.3--1.5 [Figure 5(b)]:} Faint complete, circular shell, with a small central arc detected by the {\it ROSAT} all-sky survey. This source was first proposed as an SNR by Duncan et al. (1997). The images from that paper are far poorer in angular resolution and include an unrelated close double source (centroid 12:18:01.3, --64:25:06).  

{\bf G308.4--1.4 [Figure 5(c)]:} This source is included in the list of potential SNR candidates in Whiteoak \& Green (1996), but no image was provided. The source has unusual overlapping arc morphology, making its identification uncertain. No associated {\it MSX} emission was detected above the background and there is a strong X-ray source detected by {\it ROSAT}, with insufficient angular resolution to confirm association. De Horta et al. (2013) have analysed archival radio and X-ray data to confirm the western shell is an SNR. They propose that the eastern arc is unrelated.

{\bf G310.7--5.4 [Figure 6(a)]:} Faint, complete circular shell outside the range of {\it MSX}. However, inspection of {\it IRAS} and SHASSA images do not show any measurable associated thermal emission.  

{\bf G310.9--0.3 [Figure 6(b)]:} A shell source near two known SNRs, G310.6--0.3 and G310.8--0.4, also shown in the Figure. This is not surprising as this longitude is a tangent direction to one of the spiral arms of the Galaxy (Scutum-Centaurus). The shell is lumpy and elliptical.  

{\bf G321.3--3.9 [Figure 6(c)]:} A large scalloped shell, well off the Galactic Plane. The shell is elliptical and almost complete. First proposed in Duncan et al. (1997) but their lower resolution image had included some unrelated background sources.

{\bf G322.7+0.1 [Figure 7(a)]:} Extremely faint circular shell, proposed as a potential SNR candidate in Whiteoak \& Green (1996), but no image was provided. The region contains several shell SNRs.

{\bf G322.9--0.0 [Figure 7(b)]:} Extremely faint circular shell, proposed as a potential SNR candidate in Whiteoak \& Green (1996) but no image was provided. The region contains several shell SNRs.

{\bf G323.7--1.0 [Figure 7(c)]:} Extremely faint oval shell with an HII region located a few minutes to the east. No obvious morphology match with MIR emission in the region.

{\bf G324.1+0.0 [Figure 8(a)]:} Elongated shell with two brighter arms (N and S). In a complex region, surrounded by compact HII regions and some filamentary thermal emission. Listed as a possible candidate in Whiteoak \& Green (1996) but no image was provided. 

{\bf G325.0--0.3 [Figure 8(b)]:} A small circular shell, a possible candidate suggested by Whiteoak \& Green (1996) but no image was provided. Filamentary thermal emission to the east. 

{\bf G330.7+0.1 [Figure 8(c)]:} A weak partial shell, with scalloped appearance. The southern arc is the strongest feature in the source.

{\bf G334.0--0.8 [Figure 9(a)]:} The source is a partial shell of filaments. Close to weak diffuse emission in  the MIR image, although there is no morphological match. 

{\bf G336.7--0.3 [Figure 9(b)]:} Elongated shell, possibly a composite remnant. The surrounding region is complex in both radio and MIR emission. 

{\bf G336.9--0.5 [Figure 9(c)]:} Irregular broken shell, with bright knots on the north and south rims. This is a confused region of diffuse MIR emission, but there is no morphological match to the radio features.

{\bf G345.1--0.2 [Figure 10(a)]:} Circular shell with central source, possibly a composite remnant. In a region of heavy dust extinction. Proposed as a candidate in Whiteoak \& Green (1996) but no image was provided. The source has a PMN counterpart (Wright et al. 1994) with a flux of $\sim1$ Jy, compatible with a nonthermal spectral index of about $-0.4$. The source is essentially unresolved in the PMN data (resolution $\sim5^{\prime}$) and remains a potential SNR.  

{\bf G345.2+0.2 [Figure 10(b)]:} Irregular shell with bright knots on the north and south rims. There is a small HII region to the south. The source was a possible candidate in Whiteoak \& Green (1996) but not confirmed. 

{\bf G346.2--1.0 [Figure 10(c)]:} Circular incomplete shell with variable rim intensity. The small bright source $2^{\prime}$ to the south appears to have no MIR counterpart. It may be an unrelated background source, also nonthermal.

{\bf G348.9+1.1 [Figure 11(a)]:} Irregular shell with brighter northern and southern rims. The central features have a counterpart in MIR emission. It was listed as a possible SNR candidate in Whiteoak \& Green (1996) but the  identification remains uncertain. 

{\bf G354.1+0.3 [Figure 11(b)]:} Irregular shell with brighter western rim. There ia some central emission. The source is $\sim5^{\prime}$ from a listed composite SNR G354.1+0.1, that is close to a ring of thermal emission. The region is complex and warrants further investigation.

\section{FILAMENTARY STRUCTURES IN THE ISM}
At 843 MHz the radio emission detected may be nonthermal synchrotron or thermal bremsstrahlung, but is almost always optically thin. At 8$\mu$m the emission is also mostly optically thin and is produced largely from bending modes of large PAH molecules. In comparing the two surveys we are able to see across the entire Galaxy, with the exception of the most heavily obscured and embedded sources. However, these imaging surveys provide no direct distance information so the linear sizes of the objects detected are not available.

\begin{figure*}
\begin{center}
\includegraphics[width=18cm]{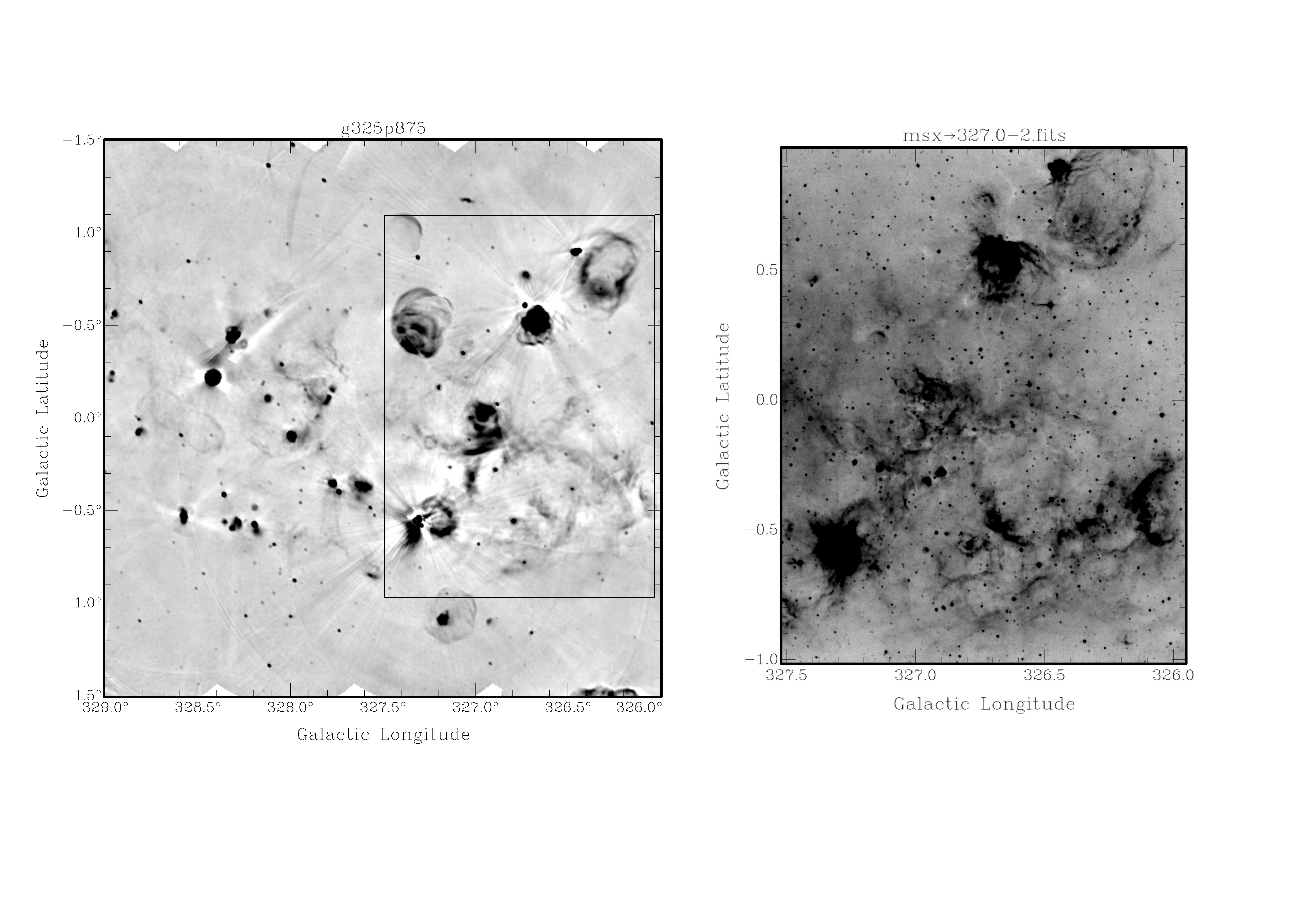}
\vspace*{-2cm}
\caption{{\it Left}: MGPS-2 image for section of the Galactic Plane at Longitude $327^{\circ}$, showing filamentary and diffuse emission together with discrete sources - HII regions, SNRS and unresolved background galaxies. The greyscale is clipped between $-4$ and $+13$ mJy beam$^{-1}$ to emphasise extended emission. {\it Right}: {\it MSX} counterpart at $8\mu$m of the SW corner of the image (shown boxed in the MGPS-2 image), demonstrating similar morphology for the HII regions and filamentary structure in the ISM. The greyscale is clipped  between 1.9$\times 10^{-6}$ and 9.3$\times 10^{-6}$ W m$^{-2}$ sr. }\label{mgps-ISM}
\end{center}
\end{figure*}

Interferometers are blind to the smoothest structures but are sensitive to rims and filaments. MGPS-2 images show much filamentary structure beyond discrete SNRs and the core sections of HII regions such as is seen around large star-forming regions like the Carina Nebula. It is unclear if these emission features are discrete objects or larger scale perturbations of the ISM. They may be the result of progressive periods of star formation and subsequent stellar winds and eventually supernova explosions. Figure 12 shows an example of this filamentary structure, much of which has juxtaposed MIR emission. There are also many examples of small arcs or filaments of apparently nonthermal emission, perhaps also SNRs or shocked outflows in the ISM, that cannot be confidently identified at present.

 Although the dominant global radio emission from a  spiral galaxy is the nonthermal diffuse synchrotron radiation that correlates so well with the far-infrared emission over a wide range of intensity (e.g. Condon 1992), the smaller scale radio emission appears to be mostly thermal. These features may be evidence of the warm ionised medium (WIM) envisaged by McKee \& Ostriker (1977) in their three-phase model of the ISM. This may also reflect the difference in sensitivity to scale between observations of the Galaxy and the integrated flux measured in extragalactic surveys. 

\begin{figure*}
\begin{center}
\includegraphics[width=13cm]{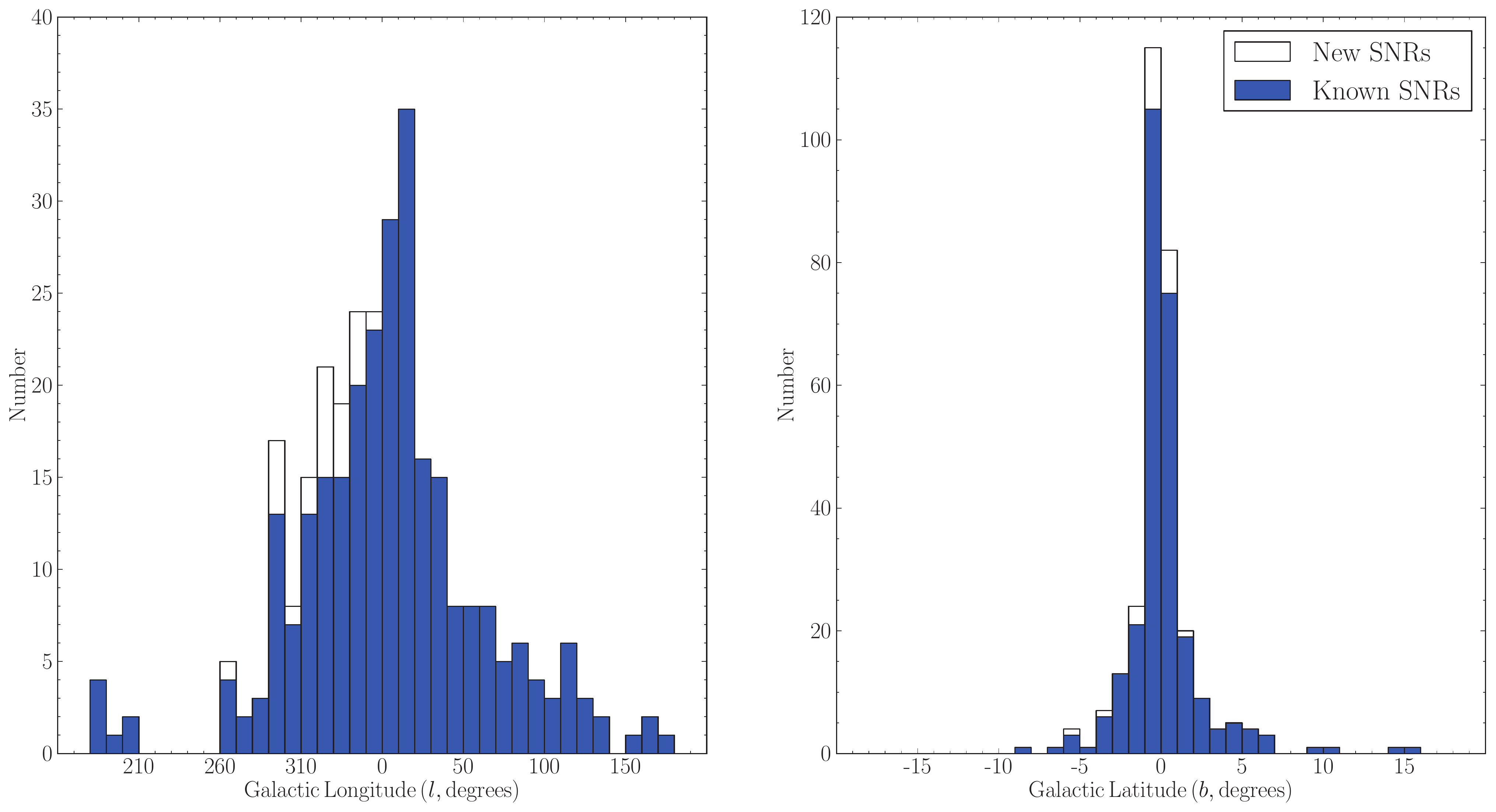}
\caption{Histograms showing the distribution of known (blue) and candidate (white) SNRs. {\it (Left):} Plot against Galactic longitude. MGPS-2 covers the range $245^{\circ} \leq l \leq +5^{\circ}$. {\it (Right):} Plot against Galactic latitude. MGPS-2 extends to $|b|\le10^{\circ}$.   }\label{snrs-plots}
\end{center}
\end{figure*}

\section{DISCUSSION}

\subsection{The Supernova Remnant Population}
The new SNR candidates have specific size and morphology criteria and it is likely that there are still more sources to be identified before there is a complete sample down to a given flux density cutoff. The candidates presented here were selected to be $\ge5^{\prime}$ in angular extent, have a shell morphology and radio emission that is anti-correlated with $8\mu$m emission from a comparable MIR survey.

Selection effects strongly affect the search for new SNRs. Green (2004) gave a comprehensive description of the limits on surface brightness and size. The current results extend the sensitivity limit of weak remnants  to below about $10^{-6}$ Jy sr$^{-1}$, equivalent to $10^{-22}$~W m$^{-2}$Hz$^{-1}$sr$^{-1}$, but the population is not yet complete to this limit. The current search size criterion of $\ge5^{\prime}$ does not change the result from Green (2004) that there is a serious lack of small diameter SNRs in current catalogues. Instrumental resolution has restricted their identification for the MGPS-2 survey but it may also be that very young SNRs still in the free expansion phase are not strongly emitting. 

The total population of SNRs to explain the energy, structures and chemical enrichment of the ISM is several times the current number of identifications (Li et al. 1991, Frail et al. 1994, Tammann et al. 1994). Current search algorithms do not have robust strategies for discriminating small angular diameter SNRs from external galaxies or for identifying very evolved SNRs that can mimic the ionisation fronts associated with thermal sources (Hollenbach \& Tielens 1997). For example, the southern bubble S145 (G308.709+0.627) discovered by Churchwell et al. (2006) has a morphology similar to many SNRs. How to identify the profile of aging radiative shells and what evidence remains after the SN shock falls below the sound speed is not clear. Past investigations of the neutral hydrogen accumulating outside the shock front do not reproduce the morphology convincingly. For example, the results for the SNR W28 presented by Velasquez et al. (2002). 

The lack of strong MIR emission at 8$\mu$m from SNRs remains a powerful discriminator (e.g. Cohen \& Green 2001). There have been detections of the IR emission from many SNRs with the IRAC and MIPS cameras of the Spitzer telescope  (e.g. Reach et al. 2006; Pinheiro Goncalves et al. 2011). However, the spectrum of a dusty SNR shows only weak emission at 8$\mu$m (see Figure 1, Pinheiro Goncalves et al. 2011), with far stronger emission at 24$\mu$m and a peak at about 50$\mu$m. The MIPS detections are only for relatively bright radio SNRs, with a median surface brightness at 1 GHz for these sources of $19\times10^{-21}$ W m$^{-2}$Hz$^{-1}$sr$^{-1}$, taken from Green (2009).  The mean value for surface brightness of our present list of candidates is $\sim10^{-21}$ W m$^{-2}$Hz$^{-1}$sr$^{-1}$. An inspection of some known weak thermal sources of comparable peak flux shows correlated {\it MSX} features for at least part of the object. Automated searches using machine learning techniques may provide new small diameter candidates but the diverse morphology of SNRs on angular scales of more than a few arcminutes remains a challenge for accurate identification. 

There is an asymmetry in the distribution and size of SNRs between the northern hemisphere (1st and 2nd quadrants)  and the south (3rd and 4th quadrants), particularly with regard to Galactic latitude. Figure 13 shows the distribution with respect to both Galactic longitude and latitude.

The different distribution in longitude may be due to the fact that the south has clear tangent directions along at least 4 spiral arms, resulting in a greater volume of the thin disk of the Galaxy being probed, whereas in the northern quadrants, the sightlines cut across smaller sections of the arms. This is presuming that most SNRs are the result of massive star collapse, and these stars are short-lived and do not move far from their nascent molecular clouds, which are distributed close to the Plane. 

The greater number of SNRs found at higher latitude in the northern hemisphere may be due to the larger number of radio telescopes undertaking surveys that are located there, particularly single dishes with lower angular resolution and the capacity to detect total fluxes. These instruments have an advantage at higher latitude. Close to the Plane, confusion is problematic and interferometers are more successful. It is hoped that the present results (with sub-arcmin resolution) will serve as finding charts for the low frequency MWA for future multiwavelength studies.   

\subsection{Filaments in the ISM}
The structures observed in MGPS-2 that are highly filamentary and of irregular morphology are mostly thermal, given their corresponding MIR emission. It may be that these features are the consequence of overlapping SNR explosions, whereby the ISM becomes a clumpy multiphase medium with a substantial fraction of WIM. The argument for this model is well explained in Chapter 39 of Draine (2011). More detailed mapping of these structures requires measurement of velocity and hence distance. Radio recombination lines at frequencies just below 1 GHz are a promising avenue, accessible in particular to new widefield instruments such as the Australia Telescope SKA Pathfinder (ASKAP; Johnston et al. 2007) and the revitalised Molonglo Telescope.   

One of the challenges is that the morphology of these features does not show the characteristic profile of a shock front, typically seen in SNR shells. Their scalloped morphology is frequently similar to the wind-blown bubbles associated with massive stars.  The present survey is not able to distinguish whether the filaments are relics of highly evolved  SNRs dissipating into the ISM or wind-blown bubbles from progenitor stars, such as are seen in the radio continuum images of thermal bubbles reported in Wendker et al. (1975), Cappa et al. (1999, 2002) and Kantharia et al. (2007).  In some cases, there is also a spatial separation between the radio and the MIR filaments, occuring around HII regions  where the size and compositon of dust grains influences emission (see example in Cohen \& Green 2001).

\section{CONCLUSIONS}

The second epoch Molonglo Galactic Plane Survey‍ (MGPS-2) undertaken with the MOST has produced a valuable archive of images that can be used as a template for future surveys of variable and transient sources as well as for current studies of the myriad of weak extended sources that the survey reveals. The survey covers 2400 deg$^2$ and includes many extragalactic sources, mostly unresolved. There is also likely to be a number of small diameter young SNRs not yet identified.

A search for SNR candidates that have a shell-like morphology with an angular diameter at least $5^{\prime}$ and an absence of measurable MIR emission, from the {\it MSX} survey data, has produced 23 new candidates. Included in the list are some candidates which were proposed earlier (Whiteoak \& Green 1996; Duncan et al. 1997) but not yet confirmed because they were observed with poorer angular resolution or were only able to be compared with the much coarser and less sensitive FIR survey from {\it IRAS}. It is planned to observe these candidates with the MWA to understand the evolution of SNRs through their low frequency properties. Using the Molonglo results will avoid potential confusion limitations with the lower frequency telescope.

The WIM contributes ~25\% to the mass fraction of the Galactic ISM and may have a very important role to play in modelling the energy budget of the Galaxy. Some of the filamentary thermal features may be the result of wind-blown bubbles from massive stars. However, some features have a braided or more diffuse character, which may be ionisation fronts or signposts of the clumpy multiphase nature of the ISM in the spiral arms. A comparison of the morphology of these structures against the HI survey of McClure-Griffiths et al. (2001) may show where future star formation is likely to occur.

\begin{acknowledgements}
The MOST  is owned and operated by the University of Sydney, with support from 
the Australian Research Council and Science Foundation within the School of
Physics. The Molonglo Observatory site staff, led by the Site Manager Duncan Campbell-Wilson, are responsible for the smooth and efficient operation of the telescope and have provided exceptional support in managing the operation of the telescope and the observations that have enabled the survey. SR  acknowledges the support of an ARC Postgraduate Award. 

\end{acknowledgements}

\end{document}